\documentclass[aps,prd,onecolumn,groupedaddress,nofootinbib,superscriptaddress]{revtex4}  
\usepackage{graphicx,mathtools,tabu}
\usepackage{epstopdf}
\usepackage{amsmath}
\usepackage{amsfonts}
\usepackage[colorlinks=true,citecolor=red,linkcolor=blue,breaklinks=true]{hyperref}
\usepackage{accents}
\usepackage[figure, table]{hypcap}
\usepackage{amssymb}
\usepackage{appendix}
\usepackage{comment}
\usepackage{bbold}
\usepackage{color}
\usepackage{slashed}
\usepackage{subfigure}
\usepackage{setspace}
\usepackage{footnote}
\usepackage{multirow}
\usepackage{braket}
\usepackage[normalem]{ulem}
\usepackage[utf8]{inputenc}
\usepackage{mathrsfs}
\usepackage{longtable}
\usepackage{enumitem}

\LTcapwidth=\textwidth

\usepackage{url}
\usepackage{wrapfig}
\usepackage{braket}

 \def\be   {\begin{equation}}  
 \def\ee   {\end{equation}}
 \def\ba   {\begin{array}}     
  \def\ea   {\end{array}}
 \def\bea  {\begin{eqnarray}}  
  \def\eea  {\end{eqnarray}}
 \def\bean {\begin{eqnarray*}}  
 \def\eean {\end{eqnarray*}}

  \def\be {\beta}

\def\to {\rightarrow}

\newcommand*{\pbar}[1]{\accentset{(-)}{#1}}

\begin{document}

 \hfill{NUHEP-TH/19-012, FERMILAB-PUB-19-508-T}
\title{On the Impact of Neutrino Decays on the Supernova Neutronization-Burst Flux}

\author{Andr\'{e} de Gouv\^{e}a}
\email{degouvea@northwestern.edu}
\affiliation{Northwestern University, Department of Physics \& Astronomy, 2145 Sheridan Road, Evanston, IL 60208, USA}
\author{Ivan Martinez-Soler}
\email{ivan.martinezsoler@northwestern.edu}
\affiliation{Northwestern University, Department of Physics \& Astronomy, 2145 Sheridan Road, Evanston, IL 60208, USA}
\affiliation{Theory Department, Fermi National Accelerator Laboratory, P.O. Box 500, Batavia, IL 60510, USA}
\affiliation{Colegio de F\'isica Fundamental e Interdisciplinaria de las Am\'ericas (COFI), 254 Norzagaray street, San Juan, Puerto Rico 00901.}
\author{Manibrata Sen}
\email{manibrata@berkeley.edu}
\affiliation{Northwestern University, Department of Physics \& Astronomy, 2145 Sheridan Road, Evanston, IL 60208, USA}
\affiliation{Department of Physics, University of California Berkeley, Berkeley, California 94720, USA}

\begin{abstract}

The discovery of non-zero neutrino masses invites one to consider
decays of heavier neutrinos into lighter ones. We investigate the
impact of two-body decays of neutrinos on the neutronization burst of
a core-collapse supernova -- the large burst of $\nu_e$ during the
first 25~ms post core bounce. In the models we consider, the $\nu_e$,
produced mainly as a $\nu_3\,(\nu_2)$ in the normal (inverted) mass
ordering, are allowed to decay to $\nu_1\,(\nu_3)$ or
$\bar{\nu}_1\,(\bar{\nu}_3)$ , and an almost massless scalar. These
decays can lead to the appearance of a neutronization peak for a
normal mass ordering or the disappearance of the same peak for the
inverted one, thereby allowing one mass ordering to mimic the
other. Simulating supernova-neutrino data at the Deep Underground
Neutrino Experiment (DUNE) and the Hyper-Kamiokande (HK) experiment,
we compute their sensitivity to the neutrino lifetime.  We find that,
if the mass ordering is known, and depending on the nature of the
Physics responsible for the neutrino decay, DUNE is sensitive to
lifetimes $\tau/m \lesssim 10^6$~{s/eV} for a galactic SN sufficiently
close-by (around 10~kpc), while HK is sensitive to lifetimes $\tau/m
\lesssim 10^7$~{s/eV}. These sensitivities are far superior to
existing limits from solar-system-bound oscillation experiments.
Finally, we demonstrate that using a combination of data from DUNE and
HK, one can, in general, distinguish between decaying Dirac neutrinos
and decaying Majorana neutrinos.

\end{abstract}

\maketitle

\section{Introduction}
\label{sec:intro}

The fact that neutrinos have nonzero masses \cite{Tanabashi:2018oca}
invites several questions related to other unknown neutrino
properties, among those the values of the neutrino lifetimes. Given
everything currently known about the neutrinos, one can affirm that
the two heavier neutrinos -- $\nu_3\,,\nu_2$ in the case of the
so-called normal mass ordering (NMO), $\nu_2\,,\nu_1$ in the case of
the so-called inverted mass ordering (IMO) -- have nonzero mass and
finite lifetime
\cite{Bahcall:1972my,Schechter:1981cv,Bahcall:1986gq,Nussinov:1987pc,Frieman:1987as,Kim:1990km,Biller:1998nc}. Assuming
no new interactions or degrees of freedom, the heavier neutrinos
decay, at the one-loop level, to either another neutrino and a photon,
$\nu_h\to\nu_l\gamma$, or to three lighter neutrinos, $\nu_h\to
\nu_l\nu_l\bar{\nu}_l$, where $h$ refers to a heavier neutrino
mass-eigenstate, $l$ to a lighter one. Under the same assumptions,
because neutrino masses are tiny, the expected lifetimes are many
orders of magnitude longer than the age of the universe.

New interactions and degrees of freedom will, of course, lead to
potentially much shorter lifetimes. One possibility, which we consider
in this paper, is the introduction of a new (almost) massless scalar,
which can couple to the neutrinos
\cite{Schechter:1981cv,PhysRevLett.45.1926,GELMINI1981411,Gelmini:1983ea,BERTOLINI1988714,SANTAMARIA1987423}. This
allows for the decay of the heavy neutrino into a lighter neutrino and
the scalar.  In these scenarios, the daughter neutrinos can play a
significant role, as we describe in
Section~\ref{sec:neutrinodecay}. The nature of the neutrinos -- Dirac
versus Majorana -- and the helicity-structure of the interaction
\cite{Kim:1990km,Beacom:2002cb,Coloma:2017zpg,Balantekin:2018ukw,Funcke:2019grs}
also play an important role since they render the daughter visible or
invisible, or render the (effective) lepton-number the same as or
different from that of the parent.

Different experimental constraints have been placed on the lifetime of
the neutrinos, ranging from bounds from terrestrial neutrinos to the
study of neutrinos produced in astrophysical sources
\cite{Berezhiani1992,Fogli:1999qt,Choubey:2000an,Lindner:2001fx,Beacom:2002cb,Joshipura:2002fb,Bandyopadhyay:2002qg,Ando:2003ie,Ando:2004qe,Beacom:2004yd,Berryman:2014qha,Picoreti:2015ika,FRIEMAN1988115,Mirizzi:2007jd,GonzalezGarcia:2008ru,Maltoni:2008jr,Baerwald:2012kc,Broggini:2012df,Dorame:2013lka,Gomes:2014yua,Abrahao:2015rba,Coloma:2017zpg,Gago:2017zzy,Choubey:2018cfz,deSalas:2018kri,Funcke:2019grs}. Roughly
speaking, the idea is simple: if neutrinos are produced in a
relatively well-characterized source and measured a certain known
distance away from the source, the fact that they make it to the
detector implies, in a very model-independent way, that the neutrinos
are not disappearing along the way. All bounds are much smaller than
the expectations of the standard-three-neutrinos-paradigm and the age
of the universe.

The best, mostly model-independent bounds on the decays of $\nu_2$ and
$\nu_1$ come from the analysis of solar neutrinos
\cite{Joshipura:2002fb,Berryman:2014qha}.  Using mostly the $^8{\rm
  B}$ solar neutrino data, one is sensitive to $\tau_2/m_2 <
10^{-3}\,{\rm s/eV}$\footnote{All neutrinos ever observed are
  ultra-relativistic in the lab frame and hence one is only sensitive
  to the ratio $\tau/m$. The fact that the neutrino masses are not
  known with any precision leads one to always quote the constraints
  on the neutrino lifetime as constraints in $\tau/m$. Historically,
  the literature prefers to quote bounds on $\tau/m$ in units of
  second per electronVolt. In ``oscillation-equivalent units,''
  \cite{Berryman:2014qha} 1~eV/s~= $6.58\times 10^{-16}$~eV$^2$.}
\cite{Berryman:2014qha}, while, using mostly the $^7{\rm Be}$ and low
energy $pp$ neutrino data \cite{Bellini:2014uqa}, one is sensitive to
$\tau_1/m_1 < 10^{-4}\,{\rm s/eV}$
\cite{Berryman:2014qha}. Considering invisible decays, the current
bounds imposed from the analysis of the SNO data, combined with other
solar neutrino experiments, results in the lifetime of $\tau_2/m_2 >
1.92\times 10^{-3}\,{\rm s/eV}$ at $90\%$ confidence
\cite{Aharmim:2018fme}. Since $|U_{e3}|^2$ is very small, solar data
are very inefficient when it comes to constraining the lifetime of
$\nu_3$. Constraints on the $\nu_3$ lifetime can be obtained from
atmospheric and beam neutrino experiments. Recent bounds on the
$\nu_3$ lifetime were estimated using atmospheric data: $\tau_3/m_3 >
10^{-10}\,{\rm s/eV}$
\cite{GonzalezGarcia:2008ru,Gomes:2014yua}. Clearly, these bounds are
much weaker than the ones on the $\nu_1$ and $\nu_2$
lifetimes. Slightly more stringent bounds
\cite{Abrahao:2015rba,Coloma:2017zpg} are expected from
next-generation neutrino oscillation experiments including the JUNO
\cite{An:2015jdp} and DUNE \cite{Acciarri:2016crz} experiments, while
stronger but model-dependent bounds -- $\tau_3/m_3 > 2.2\times
10^{-5}\,{\rm s/eV}$ -- from solar neutrinos have been recently
proposed \cite{Funcke:2019grs}.

These rather loose bounds lead one to explore baselines which are much
longer than one Astronomical Unit. Two candidates immediately qualify
for the search: (i) neutrinos from supernova SN1987A
\cite{PhysRevLett.58.1490,PhysRevLett.58.1494,1988ApJ...332..826F,Berezhiani:1989za,Kachelriess:2000qc,Farzan:2002wx},
and (ii) ultra-high-energy neutrinos from astrophysical sources,
detected at IceCube \cite{Aartsen:2014gkd,Denton:2018aml}. Constraints
from SN1987A were pursued in the literature
\cite{FRIEMAN1988115,1993APh.....1..377O}, while constraints from
IceCube -- current and future -- are plagued by uncertainties on the
neutrino flavor composition
\cite{Beacom:2002vi,Bustamante:2016ciw}. Here we concentrate on the
next galactic supernova as a probe of long lifetimes for the
neutrinos.

A core-collapse supernova (SN) emits almost $99\%$ of its binding
energy in the form of neutrinos (see \cite{Mirizzi:2015eza} for a
detailed review). During, roughly, the first 25~ms post core-bounce, a
large burst of electron neutrinos is emitted due to the
deleptonization of the core. This phase, known as the ``neutronization
burst'', is a robust prediction of all core-collapse SN
simulations. During the neutronization burst, negligible amounts --
relative to that of $\nu_e$ -- of $\overline{\nu}_e$ and
$\nu_{\mu,\,\tau},\,\overline{\nu}_{\mu,\,\tau}$ are released. This
implies that the neutronization-burst flux does not undergo
significant collective oscillations but is mostly processed by the
well-understood Mikheyev-Smirnov-Wolfenstein (MSW) resonant flavor
conversion \cite{PhysRevD.17.2369,Mikheev:1986gs} associated to the
ordinary matter density the neutrinos encounter on their way our of
the supernova. This implies that the flavor-evolution of these
neutrinos is not impacted by large uncertainties associated with
collective oscillations
\cite{Duan:2006an,Hannestad:2006nj,Fogli:2007bk,Dasgupta:2009mg,EstebanPretel:2007ec,Dasgupta:2008my,Dasgupta:2011jf,Chakraborty:2015tfa,Dasgupta:2016dbv,Izaguirre:2016gsx,Das:2017iuj,Dasgupta:2018ulw,Capozzi:2018clo}. This
renders the neutronization-burst neutrinos well suited for performing
robust tests of neutrino properties. We provide a detailed description
of these neutrinos in Section~\ref{sec:neutronization}.

A detailed analysis allowing for the decay of the heavier neutrinos
during the neutronization burst was first performed in
\cite{Ando:2004qe}. It was argued that the $\nu_e$, propagating as a
heavy mass-eigenstate $\nu_h$, could decay into a lighter
$\overline{\nu}_l$ during its flight. Considering a simple
Majoron-like model, with scalar as well as pseudo-scalar couplings,
the author studied the presence of a sharp neutronization peak due to
$\overline{\nu}_e$ in water-Cherenkov detectors. Here, we explore
different scenarios that lead to neutrinos decaying into a scalar and
a lighter neutrino, and concentrate on several physics
questions. First, we discuss whether the hypothesis that neutrinos
have a finite lifetime can hinder the ability of measurements of the
next galactic supernova neutrinos to determine the neutrino
mass-ordering. Second, assuming the mass ordering is known, we
characterize how well measurements of the neutrinos from a galactic
supernova explosion can be used to measure or place bounds on the
neutrino lifetimes. Finally, we discuss how different measurements of
the neutronization-burst neutrinos can be used to distinguish
different neutrino-decay scenarios and provide information on the
nature of the neutrino.

Our strategy is as follows. Using the results of the hydrodynamical
simulations of the Garching group for a $25\,M_\odot$ progenitor model
\cite{Garching}, we study the neutronization-burst flux in the
presence of a neutrino two-body decay, $\nu_i \to
\pbar{\nu}_j\,\varphi$, where $\varphi$ is a massless scalar; the
details are spelled out in Section~\ref{sec:decaysupernova}. We
analyze the signatures of this decay in the upcoming DUNE experiment
\cite{Acciarri:2016crz}, and the Hyper-Kamiokande (HK) experiment
\cite{Abe:2018uyc}. We find that for a typical supernova located at a
distance of $10\,{\rm kpc}$, these experiments are sensitive to
lifetimes $\tau/m\lesssim 10^5-10^7\,{\rm s/eV}$, which are much
longer than what is currently constrained by solar and long-baseline
neutrino data. Furthermore, since DUNE is mainly sensitive to the
$\nu_e$ spectrum, while HK is mostly sensitive to the $\bar{\nu}_e$
spectrum, a combination of these two experiments may be able to inform
the Majorana versus Dirac question. Details of our simulation are
described in Section~\ref{sec:experiments} while our results are
discussed in Section~\ref{sec:results}. We offer some concluding
remarks in Section~\ref{sec:conclusions}.

\section{Neutrino Decay into a Scalar and a Lighter Neutrino}
\label{sec:neutrinodecay}

In this section, we discuss different scenarios that allow for the
two-body decay of the neutrino mass-eigenstates, where the heaviest
neutrino $(\nu_h)$ decays to a lighter neutrino $(\nu_l)$, and a
scalar field with negligible mass. We consider the possibility that
the neutrinos are massive Majorana or Dirac fermions. For
concreteness, we will concentrate on the NMO and the scenario where
the heaviest neutrino $(h=3)$ decays to the lightest neutrino $(l=1)$,
unless otherwise noted. We will also ignore the mass of the scalars
and the lighter neutrino. It should be clear that all the results to
be derived here apply to, for example, the IMO and the scenario where
the heaviest neutrino $(h=2)$ decays to the lightest neutrino
$(l=3)$. In the next sections, we will also make reference to this
decay.

\subsection{Dirac neutrinos}

We augment the standard model Lagrangian by postulating the existence of new scalar fields. For simplicity and in order to avoid further problems, we assume the new scalar fields are singlets of the standard model gauge group and only concentrate on the lowest-dimensional couplings of these to the standard model neutrinos. If the neutrinos are Dirac fermions, lepton number is a conserved global symmetry and we can classify the new scalars according to how they transform under $U(1)_L$ -- lepton-number symmetry. We will consider two simple cases: lepton-number zero scalars and lepton-number two scalars. 

A scalar field $\varphi_0$ with zero lepton number can couple to neutrinos only at the dimension-five level:
\begin{equation}
 \mathcal{L}_{Dir}\supset  \frac{\tilde{g}_{ij}}{\Lambda} (L_i H)\nu^c_j \varphi_o+ {\rm h.c.}\, \supset g_{ij} \nu_i \nu^c_j \varphi_0 + {\rm h.c.}\,,
 \label{eq:DiracL}
\end{equation}
where $g_{ij}=\tilde{g}_{ij}v/\Lambda$, $\Lambda$ is the effective scale of the operator and $v$ is the vacuum expectation value of the neutral component of the Higgs field. Here and throughout we express all fermions as left-handed Weyl fields; $\nu^c_i$ are the left-handed antineutrino fields while $L_i$ are the standard model lepton doublets. The indices $i,j=1,2,3$ run over the different neutrino mass eigenstates.  
 
In the laboratory frame, the decay width of such a process is 
\begin{equation}
 \Gamma= \frac{g^2 m_3^2}{64\pi E_3}\,,
 \label{eq:width}
\end{equation}
where $g^2 = |g_{13}|^2+|g_{31}|^2$, $m_3$ is the mass of $\nu_3$, and $E_3$ its energy.
This decay-width is related to the lifetime, $\tau_3$  in the $\nu_3$ rest-frame: $\tau_3=m_3/(E_3\,\Gamma)$. 

We will be interested in $\nu_3$'s produced via the weak interactions under conditions where, in the laboratory frame, $E_3\gg m_3$ so the $\nu_3$ beam will be, effectively, left-handed. After the $\nu_3$ decay, the daughter $\nu_1$ may be either left-handed or right-handed. The energy distributions of the daughter neutrinos in the laboratory frame are
\begin{eqnarray}
 \psi_{\rm h.c.}(E_3,E_1)\equiv\frac{1}{\Gamma}\,\frac{d\,\Gamma}{d\,E_1}& \propto&\frac{2 E_1}{E_3^2}\qquad\qquad\qquad\,\,{\rm for\,\,} \nu_{3_L} \to \nu_{1_L} + \varphi_0\,\,, \nonumber \\
 \psi_{\rm h.f.}(E_3,E_1)\equiv\frac{1}{\Gamma}\,\frac{d\,\Gamma}{d\,E_1}& \propto&\frac{2}{E_3}\left(1-\frac{E_1}{E_3}\right)\qquad{\rm for\,\,} \nu_{3_L} \to \nu_{1_R} + \varphi_0\,\,, 
 \label{eq:energydist}
 \end{eqnarray}
for the helicity-conserving (h.c.) and helicity-flipping (h.f.) cases respectively. Here `L' and `R' stand for left and right helicity of the neutrinos. In the helicity-conserving case, the $\nu_1$ spectrum is harder. This is easy to understand from simple angular momentum conservation considerations. In the $\nu_3$ rest frame, the daughter neutrinos that share the same polarization as the parent are emitted preferentially in the direction of the polarization of the parent, while those with the opposite polarization are preferentially emitted in the opposite direction. In the reference frame where the parent is boosted, the daughters with the same helicity as the parent are emitted predominantly in the forward direction (higher energy), while the opposite-helicity daughters are emitted preferentially backwards (lower energy). 
 
Ignoring effects proportional to the mass of the daughter neutrino
($m_1$), the helicity-flipping channel produces neutrinos of the
``wrong'' helicity, i.e, right-handed neutrinos, which are effectively
(and very safely) invisible to any detector. The relative weights of
the helicity-flipping and helicity-conserving channels depend on the
relative magnitudes of $g_{13}$ and $g_{31}$, related to how much the
new interactions violate parity. In maximally parity violating
scenarios, the $\nu_3$ decay may be completely invisible or completely
visible. For concreteness, when we discuss this $\varphi_0$-model
quantitatively, we will always impose the constraint $g_{ij}=g_{ji}$,
in such a way that the physics that describes neutrino decay is parity
conserving. In this case, half of the neutrino daughters from the
decay of a heavier neutrino will be of the ``correct'' helicity while
the other half will be of the ``wrong'' helicity.

A scalar field $\varphi_2$ with lepton-number two can couple to neutrinos only at dimension-four and dimension-six  \cite{Berryman:2018ogk}:
\begin{equation}
 \mathcal{L}_{Dir}\supset  \frac{y_{ij}}{2}\nu^c_i \nu^c_j \varphi_2 + \frac{\tilde h_{ij}}{2\Lambda^2}\,(L_i H)(L_j H)\varphi_2^*  + {\rm h.c}\,,
 \label{eq:DiracL2}
\end{equation}
where, after spontaneous symmetry breaking, the second term includes $h_{ij}\nu_i\nu_j\varphi$, $h_{\ij}=\tilde{h}_{\ij}v^2/\Lambda^2$. Here, $\Lambda$ is the effective scale of the dimension-six operator and $y_{ij}=y_{ij}$ and $h_{ij}=h_{ji}$.

Since $\varphi_2$ carries lepton-number, Eq.~(\ref{eq:DiracL2}) mediates the following decay: $\nu_3\to\bar{\nu_1}\varphi_2$; the neutrino lepton-number changes by two units. Assuming $y_{ij}\gg h_{ij}$, (or vice-versa), the interactions in Eq.~(\ref{eq:DiracL2}) are strongly parity violating and the decays yield strongly polarized daughters. If $y_{ij}\gg h_{ij}$, all daughters will be of the ``wrong-helicity'' type and hence invisible assuming an ultra-relativistic parent beam produced via the weak interactions (i.e., in the decay $\nu_{3L}\to\bar{\nu}_1\varphi_2$ all $\bar{\nu}_1$ will be left-handed and hence invisible in the limit $m_1/E_1\to 0$). Instead, If $h_{ij}\gg y_{ij}$, all the daughters will be of the ``correct-helicity'' and hence expected to interact in a neutrino detector (i.e., in the decay $\nu_{3L}\to\bar{\nu}_1\varphi_2$ all $\bar{\nu}_1$ will be right-handed and hence visible in the limit $m_1\to 0$). In both cases, the energy distributions are as prescribed in Eqs.~(\ref{eq:energydist}), where the invisible decays ($y_{ij}\gg h_{ij}$) follow the helicity-conserving equation, while the visible decays $h_{ij}\gg y_{ij}$ follow the helicity-flipping one. 

\subsection{Majorana neutrinos}

If the neutrinos are Majorana fermions, lepton number is explicitly broken and there is no need for light right-handed-neutrino degrees of freedom. If a new gauge-singlet scalar $\varphi$ is added to the standard model particle content,  the most minimal interaction that leads to the neutrino decays of interest is dimension-six:
\begin{equation}
 \mathcal{L}_{Maj}\supset  \frac{\tilde{f}_{ij}}{2\Lambda^2}\, (L_i H)(L_j H)\varphi + {\rm h.c.}\, \supset \frac{f_{ij}}{2} (\nu_L)_i (\nu_L)_j \varphi + {\rm h.c.}\,,
 \label{eq:MajL}
\end{equation}
where $\tilde{f}_{ij}=\tilde{f}_{ji}$, $f_{ij}=\tilde{f}_{ij}v^2/\Lambda^2$ and $\Lambda$ is the effective scale of the dimension-six operator. Here it is not meaningful to assign lepton-number to $\varphi$. 

Here, in the laboratory frame, the decay width for $\nu_3\to\nu_1+\varphi$ is 
\begin{equation}
 \Gamma= 2\times \frac{f^2 m_3^2}{64\pi E_3}\,,
 \label{eq:widthMaj}
\end{equation}
where $f\equiv f_{13}$, $m_3$ is the mass of $\nu_3$, and $E_3$ its energy. This decay-width is related to the lifetime $\tau_3$  in the $\nu_3$ rest-frame: $\tau_3=m_3/(E_3\,\Gamma)$. For Majorana neutrinos, the decay rates are twice as large as that of Dirac neutrinos (Eq.\,(\ref{eq:width})) since one is compelled to include both the ``neutrino'' and ``antineutrino'' final states.

In the limit $m_1/E_1\to 0$, it is convenient to identify the two different helicity states of $\nu_1$ as the ``neutrino'' ($\nu_{1L}$) and the ``antineutrino'' ($\bar{\nu}_{1R}$) where the charged-current scattering of the $\nu_1$ state leads to the production of negatively charged leptons while that of $\bar{\nu}_1$ leads to positively charged leptons. In this case, Eq.~(\ref{eq:MajL}) mediates both $\nu_{3L} \to \nu_{1L} + \varphi$, and $\nu_{3L} \to \bar{\nu}_{1R} + \varphi$, both of which are visible modes.  Furthermore, as long as the $\nu_1$ mass is small enough, the energy distributions are also given by Eq.\,(\ref{eq:energydist}) \cite{Kim:1990km,Beacom:2002cb}; the ``neutrino'' final state is helicity-conserving, while the ``antineutrino'' final state is helicity-flipping. The relative-branching ratios, on the other hand, are the same. 

\subsection{Majorana versus Dirac}

In general, Majorana and Dirac neutrino decays are distinguishable as long as one can measure the daughter neutrinos. In the $\varphi_0$-model (Eq.~(\ref{eq:DiracL})), heavier neutrinos decay to lighter visible neutrinos (or invisible neutrinos) while in the $\varphi_2$-model (Eq.~(\ref{eq:DiracL2})), heavier neutrinos decay exclusively into visible antineutrinos ($h_{ij}\gg y_{ij}$) or invisible antineutrinos $y_{ij}\gg h_{ij}$). The two models are distinguishable, in principle, if a nonzero fraction of the decays is visible. 

Instead, in the Majorana case, heavier neutrinos decay into both visible lighter ``neutrinos'' and ``antineutrinos'', and the branching ratios are the same.  This means that if heavy neutrinos are produced in some far away source via weak-interaction processes involving, say, negatively charged leptons (for example the deleptonization process of interest here, $e^-+p\to n+\nu$) and only one new light scalar exists, Dirac neutrinos will decay into either neutrinos or antineutrinos. Majorana neutrinos, on the other hand, will decay into both ``neutrinos'' and ``antineutrinos.''

The result above can be clouded by postulating more light degrees of freedom. If, for example, one combines the $\varphi_0$-model with the $\varphi_2$-model, it is possible to choose couplings in such a way that a Dirac $\nu_3$ decays both into $\nu_1+\varphi_0$ and $\bar{\nu}_{1}+\varphi_2$ with the same branching ratio. If this were the case, the Dirac neutrino case would perfectly mimic the Majorana one. On the other hand, if a new light fermionic degree of freedom exists, it is possible to write down an interaction like Eq.~(\ref{eq:DiracL}) for Majorana neutrinos, where the left-handed antineutrino field is replaced by the new, sterile fermion. In this case, it is possible to have the Majorana neutrino case perfectly mimic the Dirac one. We consider both of these scenarios rather finely tuned and will ignore them henceforth. 

Note that Eqs.~(\ref{eq:energydist}) are, strictly speaking, only true if there exists a large hierarchy between the masses of the neutrinos, i.e., $m_3 \gg m_1$ \cite{Beacom:2002cb}. If the neutrino masses are quasi-degenerate such that $m_3\sim m_1$ (keeping $\Delta m_{31}^2$ within observational bounds), then the helicity-flipping channel is suppressed. However, this does not impact our results for the $\varphi_0$-model, as the helicity-flipping channel is not detected. On the other hand, the helicity-flipping channel plays an important role for the $\varphi_2$ model, or if the neutrinos are Majorana. As a result, the conclusions inferred in these cases will have to be revised for the quasi-degenerate scenario. We will consider the neutrino masses to be hierarchical, and not focus on the quasi-degenerate limit in the rest of the paper.

\subsection{Order of Magnitude Considerations}

Before discussing the supernova explosion as a source of neutrinos and how the propagation of supernova neutrinos is impacted by -- and informs -- the neutrino lifetime, it is useful to estimate the naive sensitivity of supernova neutrinos to the neutrino lifetime. Lifetime effects are visible when $\Gamma\times L\gtrsim 1$, where $L$ is the neutrino propagation distance. In the case of the $\varphi_0$-model with Dirac neutrinos, this translates into
\begin{equation}
|g|\gtrsim  2.3\times 10^{-9}\left(\frac{E_3}{10~\rm MeV}\right)^{1/2}\left(\frac{10~\rm kpc}{L}\right)^{1/2}\,\left(\frac{0.5\,\rm eV}{m_3}\right).
\end{equation}
Similar values for the relevant couplings are obtained for the other models discussed in this section. This implies the systems under investigation here are sensitive to very weakly-coupled new degrees of freedom. This, in turn, implies that virtually all laboratory probes -- and many early universe probes -- will not be as sensitive to this type of new physics as the observables we are discussing here. There are a few exceptions; we comment on these in the concluding section. 

The constraint $\Gamma\times L\gtrsim 1$ can be re-expressed as $\tau/m \lesssim L/E$\footnote{$L(m/E)$ is the baseline in the neutrino rest-frame.} or 
\begin{equation}
\frac{\tau}{m}\lesssim 10^5~{\rm s/eV}\left(\frac{L}{10~\rm kpc}\right)\left(\frac{10~\rm MeV}{E}\right).
\label{eq:estimate}
\end{equation}
This implies that supernova neutrinos produced 10~kpc away are sensitive to neutrino lifetimes that approach a few days for neutrino masses of order 1~eV.

\section{Flavor evolution during the neutronization burst}
\label{sec:neutronization}

Modern day simulations of core-collapse supernovae including a detailed treatment of the neutrino transport problem agree on the presence of the neutronization-burst phase that occurs typically for around 25~ms, immediately after core-bounce \cite{Janka:2006fh}. The core-collapse shock, launched as a result of the stiffening of the core, results in the dissociation of the surrounding nuclei into individual neutrons and protons. The newly formed protons capture electrons, producing a large burst of $\nu_e$, thereby leading to the prompt deleptonization of the core. As the shock reaches relatively lower densities, these $\nu_e$ can stream out, leading to the neutronization burst. During this period, the other types of neutrinos -- $\bar\nu_e$, as well as the other flavors $(\nu_\mu,\,\bar\nu_\mu,\,\nu_\tau,\,\bar\nu_\tau)$ -- are also emitted, but the associated fluxes are very small when compared to that of the $\nu_e$.
  
A SN acts like an approximate blackbody, and cools by the emission of neutrinos. Keeping this in mind, SN simulations typically fit the neutrino spectra with the so-called ``alpha-fit''~\cite{Keil:2002in}:
 \begin{equation}
  \phi(E)=\frac{1}{\langle E_\nu \rangle}\frac{(\alpha+1)^{(\alpha+1)}}{\Gamma (\alpha+1)}\left(\frac{E}{\langle E_\nu \rangle}\right)^{\alpha}{\rm exp}\left[-(\alpha+1) \frac{E}{\langle E_\nu \rangle}\right]\,,
  \label{spectra_ch1}
 \end{equation}
where $\phi(E)$ is normalized to unity. Here, $\langle E_\nu\rangle$ denotes the average energy of the neutrinos, $\Gamma(z)$ is the Euler gamma function, and $\alpha$ is a dimensionless parameter, referred to as the pinching parameter. It is related to the width of the spectrum and satisfies 
\begin{equation}
 \frac{1}{1+\alpha}=\frac{\langle E^2\rangle-\langle E\rangle^2}{\langle E\rangle^2}\,.
\label{eq:alpha}
\end{equation}
These spectral parameters are time dependent and depend on the different neutrino emission phases, the particulars of the simulation, such as the treatment of neutrino transport, and inputs from nuclear physics. 

\begin{figure}[!t]
\includegraphics[width=0.45\textwidth]{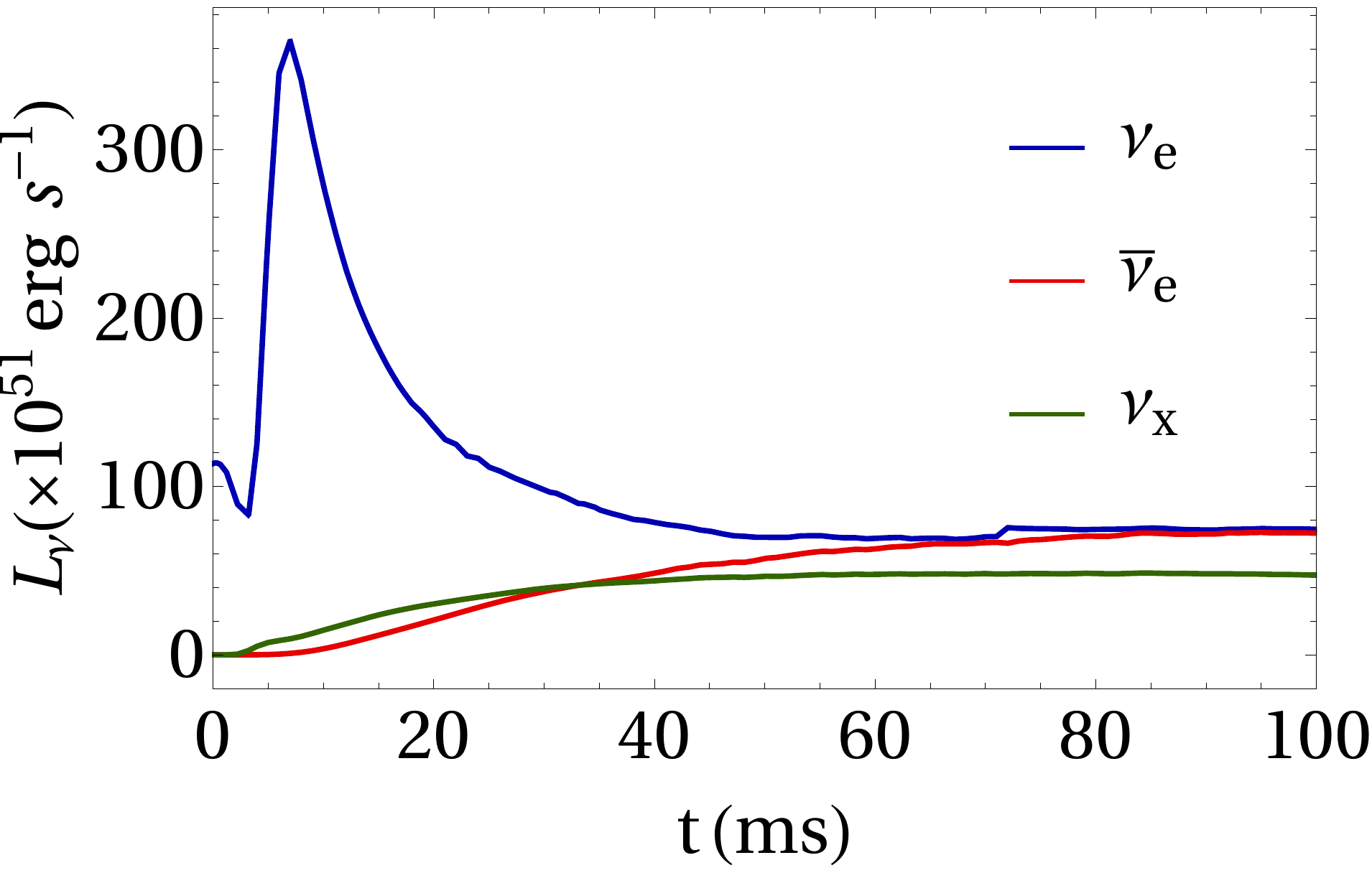}\,\includegraphics[width=0.45\textwidth]{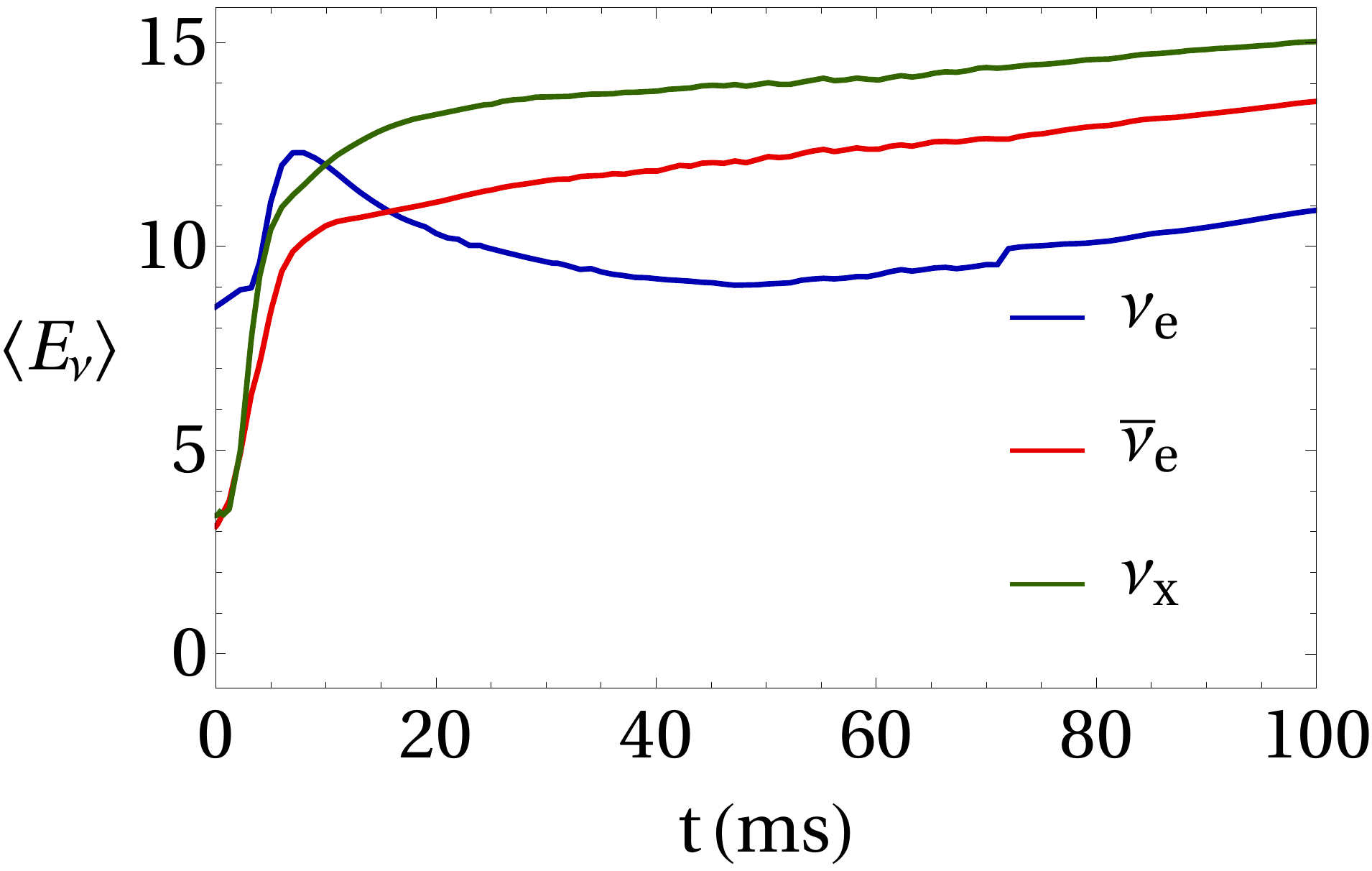}\\\vspace{0.5cm}\includegraphics[width=0.45\textwidth, height=0.3\textwidth]{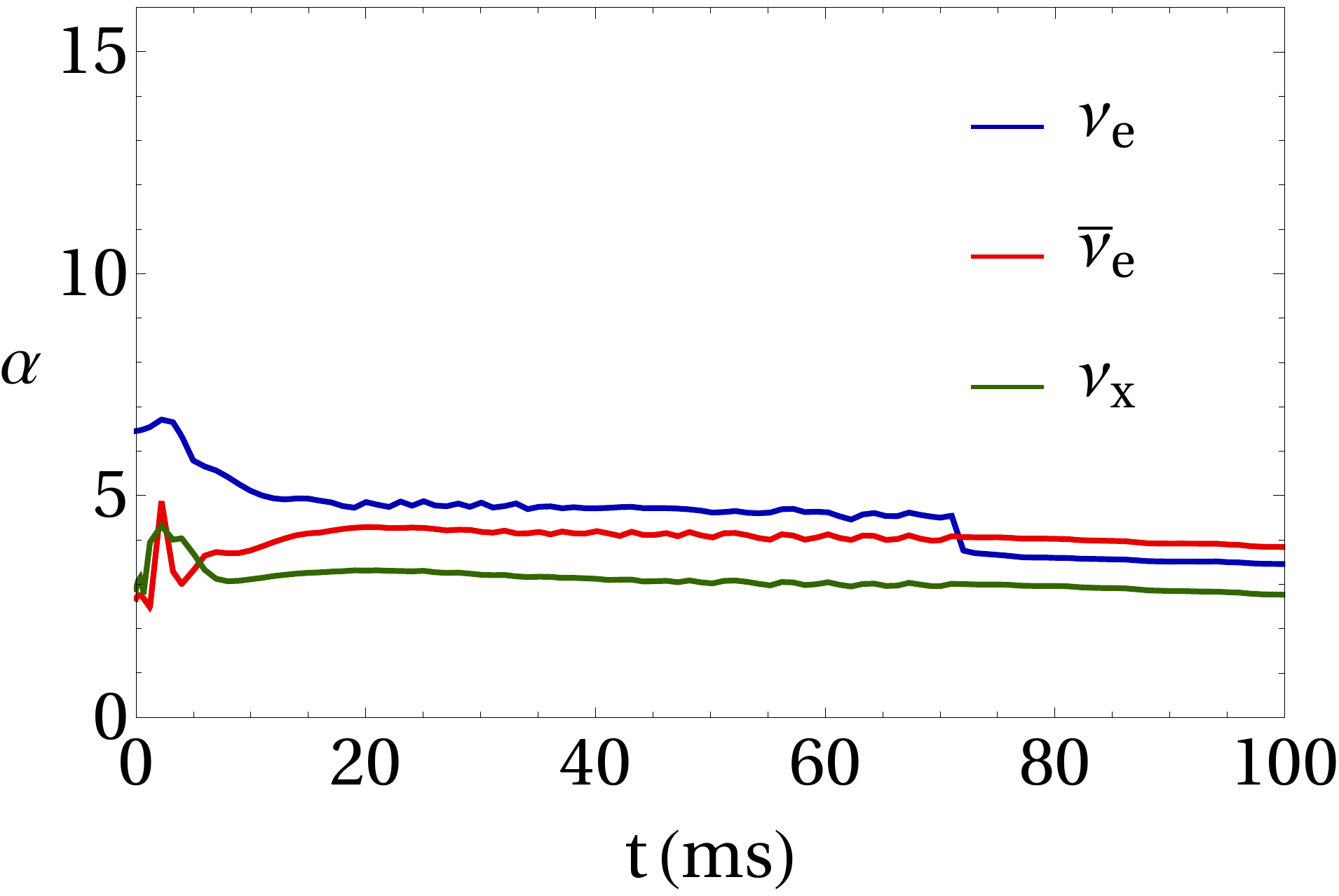}\,
\caption{Luminosities, the average neutrino energies and pinching parameter $\alpha$ -- see Eq.~(\ref{eq:alpha}) -- of the neutrinos emitted during the neutronization burst phase -- first 25~ms or so -- and slightly beyond. These results come from hydrodynamical simulations of the Garching group for a $25\,M_\odot$ progenitor model \cite{Garching}.}
\label{fig:simulations}
\end{figure}

Simulations provide the unoscillated neutrino distribution as a function of time, in the form of neutrino luminosities $L_\nu$. The neutrino distribution in time and energy is  \cite{Serpico:2011ir}
\begin{equation}
 \Phi_\nu(E,t)=\frac{L_\nu (t)}{\langle E_\nu \rangle}\,\phi(E)\,.
\end{equation}
Fig.\,\ref{fig:simulations} depicts the time evolution of the neutrino luminosities, their average energies, as well as the pinching parameters. These were obtained from a 1D (spherically symmetric) simulation of a $25\,M_\odot$ progenitor by the Garching group \cite{Garching}.

The neutronization signal is a quite robust feature of all simulations. The shape and height of the peak depend on the progenitor masses, as well as the nature of simulations (for a comparison of different simulations, see \cite{Pan:2018vkx,OConnor:2018sti}). Such details will not affect the results discussed here and in the next sections. Furthermore, since there is a large hierarchy between the fluxes of the different neutrino flavors, collective oscillations are suppressed. This implies that the calculation of the flavor-evolution of neutrinos produced during the neutronization burst is straightforward and does not suffer from the uncertainties associated with the non-linear effects due to neutrino self-interactions. 

The $\nu_e$ are produced deep inside the supernova where the local mater densities are very high. Standard matter effects imply that the $\nu_e$ is very well aligned with the instantaneous Hamiltonian eigenstate associated with the largest instantaneous Hamiltonian eigenvalue at production. The subsequent evolution is such that the different eigenstates can be treated as incoherent, and the position-dependency of the matter density, together with the known values of the neutrino mass-squared differences, indicate that the adiabatic approximation works very well. Taking all of this into account, the neutronization-burst neutrino flux at the Earth, ignoring the small $\nu_x$ initial population, is well approximated by \cite{Dighe:1999bi}
\begin{equation}
f_{\nu_e}(E,t) = \frac{1}{4\pi R^2}|{\rm U}_{eh}|^2 \Phi_{\nu_e}(E,t)
\label{eq:NueEarth}
\end{equation}
where $h$ stands for the heaviest neutrino mass eigenstate, $h=3$ for the NMO, $h=2$ for the IMO. $U_{eh}$ is the relevant element of the leptonic mixing-matrix and $R$ is the distance between the Earth and the SN. It is straightforward to include the contribution of the $\nu_x$ initial population. Since $\nu_e$ produced in the neutronization burst exit the supernova as $\nu_h$, $\nu_{\mu}$ and $\nu_{\tau}$ exit the supernova as mixtures of the two lightest mass eigenstates. Since the initial fluxes of $\nu_{\mu}$ and $\nu_{\tau}$ are roughly the same, their contribution to the measured $\nu_e$ flux at the detector is proportional to $(1-|{\rm U}_{eh}|^2) \Phi_{\nu_x}(E,t)$. For the sake of discussion, we often ignore the $\nu_x$ contribution. Nonetheless, in all numerical computations and results presented, their contribution is included. 

\begin{figure}[!t]
  \includegraphics[width=0.45\textwidth]{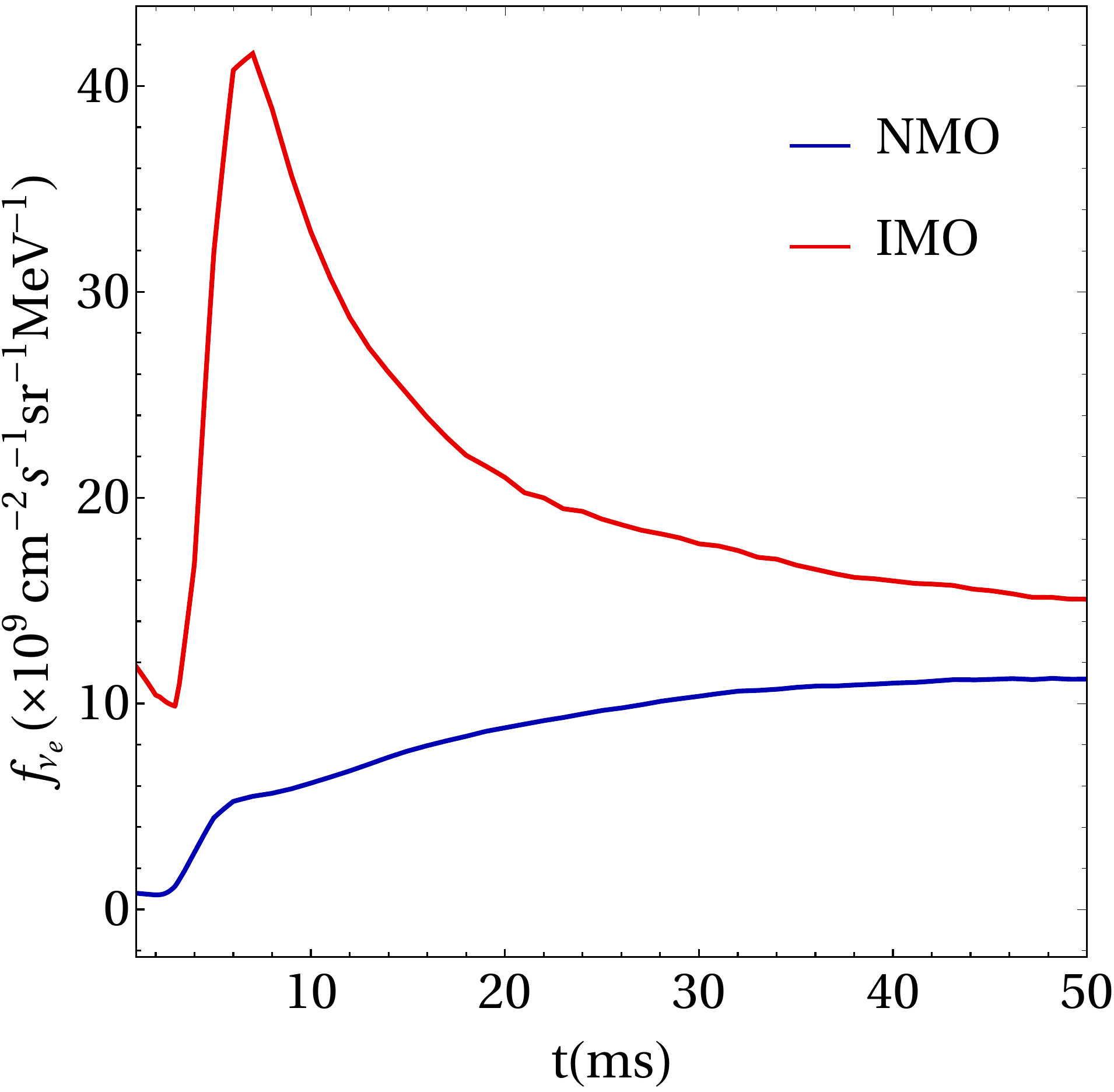}\,\qquad\quad\includegraphics[width=0.45\textwidth, height=0.44\textwidth]{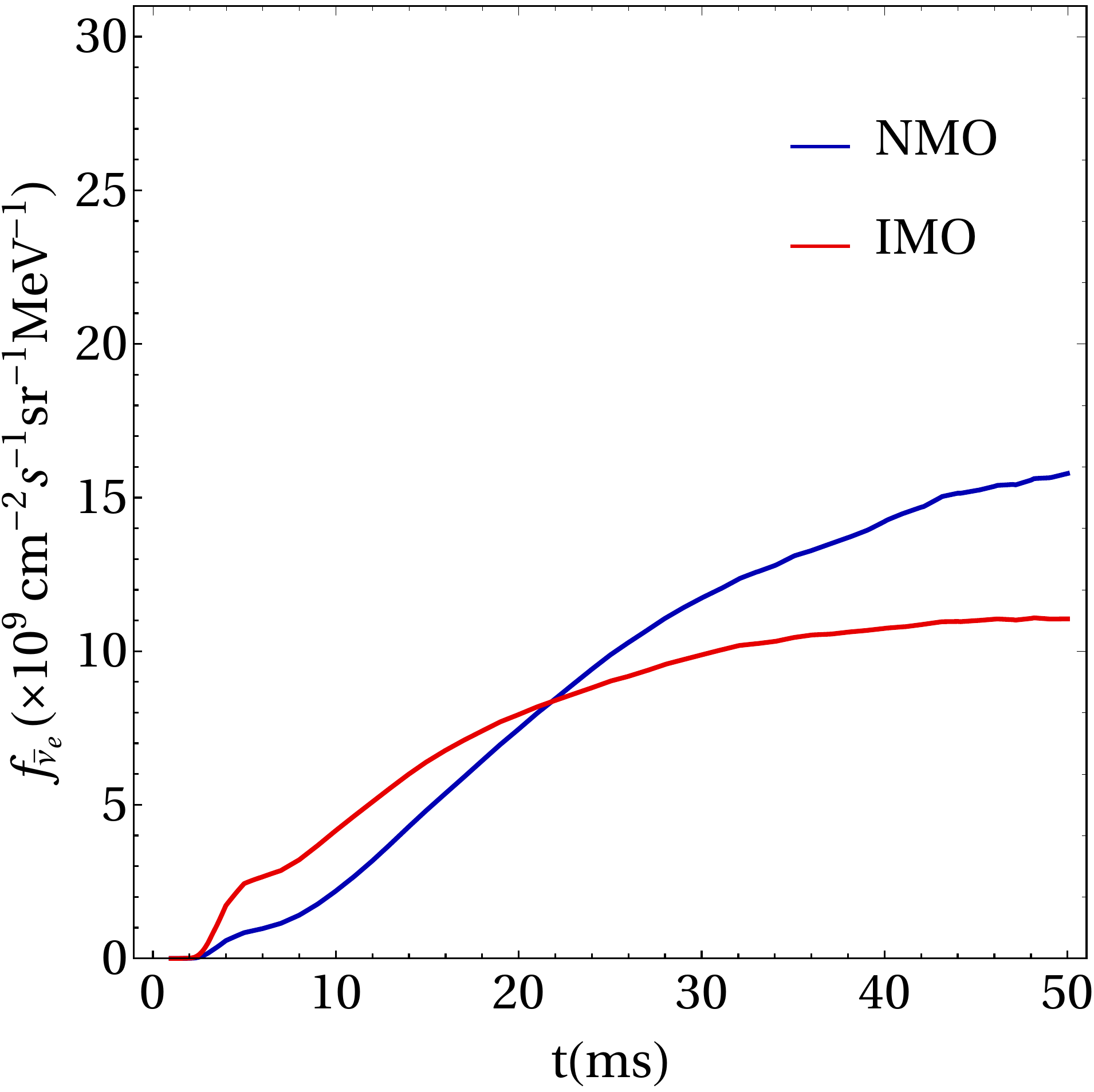}
\caption{Expected differential flux of electron neutrinos $\nu_e$ (left) and antineutrinos $\bar{\nu}_e$ (right) at the surface of the Earth as a function of time-after-bounce, for a supernova explosion 10~kpc away, for the NMO (blue) and IMO (red) and $E_{\nu}=12$~MeV. The values of the neutrino oscillation parameters are the best-fit ones tabulated in \cite{Esteban:2018azc}. The neutronization burst exists roughly for the first 25~ms post core bounce. }
\label{fig:massordering}
\end{figure}

The $\nu_e$ flux detected at the Earth is sensitive to the neutrino mass ordering. If the mass ordering is normal, a $\nu_e$ produced in the SN comes out, mostly, as a $\nu_3$, whereas if the ordering is inverted, it comes out mostly as a $\nu_2$. Since the production of other flavors is mostly negligible during this epoch, this implies that for the normal mass ordering, the $\nu_e$ flux will have a relative suppression factor $\sim |{\rm U}_{e3}|^2/ |{\rm U}_{e2}|^2 \simeq 0.1 $. This allows for the determination of the neutrino mass ordering from the neutronization-burst phase \cite{Dighe:1999bi}. This is illustrated in the left panel of Fig.\,\ref{fig:massordering}, where the $\nu_e$ fluxes to be detected at the Earth for the two mass orderings are depicted. The presence of a peak in the time spectrum is characteristic of the inverted mass ordering, whereas its absence indicates the normal ordering. 

Antineutrinos can be treated in the same way. In this case, however, since the sign of the matter potential is inverted for antineutrinos, a $\bar{\nu}_e$ produced in the center of the supernova during the neutronization burst will exit the supernova as $\bar{\nu}_l$, the lightest antineutrino mass eigenstate. In this case, $\bar\nu_{\mu,\tau}$ will exit the explosion as mixtures of the two heaviest mass-eingenstates. The $\bar{\nu}_e$ flux on Earth is depicted in the right panel of Fig.\,\ref{fig:massordering}. Here, the distinction between the two mass-orderings is a lot less pronounced. Nonetheless, it has been argued in the literature that the rise time of the $\bar{\nu}_e$-flux can be exploited in very high statistics experiments, like IceCube \cite{Dighe:2003be,Serpico:2011ir}.

\section{Impact of neutrino decay on the neutronization-burst flux}
\label{sec:decaysupernova}

 In this section, we outline the formalism for simulating the $\nu_e$ and the $\bar{\nu}_e$ flux during the neutronization-burst epoch, incorporating both neutrino propagation and decay. The impact of neutrino decay on its propagation is given by solving a transfer equation, which takes into account the decay of the heaviest-state population, as well as the repopulation of the lighter states \cite{Ando:2004qe}. The problem is simplified if we assume that the neutrinos do not decay within the SN envelope. Therefore, the neutronization-burst flux arrives unchanged at the surface of the SN, and decay on their way to the Earth. This is a good approximation as long as $\tau/m\gg 10^{-7}$s/eV. 
 
 The flux of a Dirac neutrino mass eigenstate $i$ arriving at the Earth satisfies, for the $\varphi_0$-model (Eq.~(\ref{eq:DiracL})),
\begin{equation}
 \frac{d }{dr}f_{\nu_i}(E,r)=-\Gamma_{\nu_{i}\to\nu_{j}}(E) f_{\nu_i}(E,r) + \int_{E}^{\infty}\,dE'\left[\psi_{\rm h.c}(E',E) \Gamma_{\nu_{j}\to\nu_{i}} (E') f_{\nu_j}(E',r) 
  \right],
 \label{eq:transfereq0}
\end{equation}
where $\psi_{h.c}, \psi_{h.f}$ are defined in Eq.~(\ref{eq:energydist}) and discussed in Section~\ref{sec:neutrinodecay}. We have considered the fact that the daughter-neutrinos produced with the ``wrong'' helicity are effectively invisible. The first term accounts for the disappearance of the parent-neutrinos while the second one accounts for the appearance of the daughters. The same equation, with $\nu\leftrightarrow\bar{\nu}$, holds for antineutrinos.

On the other hand, for the $\varphi_2$-model (Eq.~(\ref{eq:DiracL2})),
\begin{eqnarray}
 \frac{d }{dr}f_{\nu_i}(E,r) =-\Gamma_{\nu_{i}\to\overline{\nu}_{j}}(E) f_{\nu_i}(E,r) 
  + \int_{E}^{\infty}\,dE'\left[\psi_{\rm h.f}(E',E) \Gamma_{\overline{\nu}_{j}\to\nu_{i}} (E') f_{\bar{\nu}_j}(E',r)   \right]. \nonumber \\
   \frac{d }{dr}f_{\bar{\nu}_i}(E,r) =-\Gamma_{\bar{\nu}_{i}\to\nu_{j}}(E) f_{\bar{\nu}_i}(E,r) 
  + \int_{E}^{\infty}\,dE'\left[\psi_{\rm h.f}(E',E) \Gamma_{\nu_{j}\to\bar{\nu}_{i}} (E') f_{\nu_j}(E',r)   \right]. 
 \label{eq:transfereq2}
\end{eqnarray}
We have considered the fact that the daughter-(anti)neutrinos produced with the ``same'' helicity are effectively invisible. Here, all neutrino decays lead to antineutrino daughters, hence all daughters are ``lost.'' On the other hand, the neutrino flux is replenished with daughters from parent-antineutrinos. For the neutronization burst, this is mostly irrelevant for the $\nu_e$ detection on Earth but is very important for the $\bar{\nu}_e$ signal. 

In the case of Majorana neutrinos, taking advantage of the fact that the neutrino masses are much smaller than the neutrino energies in the lab frame, we can also define a ``neutrino'' and ``antineutrino'' flux. In this case, assuming the decay scenario associated to Eq.~(\ref{eq:MajL}),
\begin{eqnarray}
 \frac{d }{dr}f_{\nu_i}(E,r)&=&-\left( \Gamma_{\nu_{i}\to\nu_{j}}(E)+\Gamma_{\nu_{i}\to\overline{\nu}_{j}}(E)\right) f_{\nu_i}(E,r) \nonumber\\ 
 &+& \int_{E}^{\infty}\,dE'\left[\psi_{\rm h.c}(E',E) \Gamma_{\nu_{j}\to\nu_{i}} (E') f_{\nu_j}(E',r) + \psi_{\rm h.f}(E',E) \Gamma_{\overline{\nu}_{j}\to\nu_{i}} (E') f_{\bar{\nu}_j}(E',r)   \right]
 \label{eq:transfereq}
\end{eqnarray}
The same equation holds also for antineutrinos with $\nu\leftrightarrow\bar{\nu}$.

Here, we focus on the situation where the heaviest mass eigenstate decays to the lightest one: in the NMO (IMO), the $\nu_3 \,(\nu_2)$ decays to a $\nu_1\,(\nu_3)$, leaving the intermediate mass-eigenstate -- $\nu_2$ ($\nu_1$)--  unchanged. There is no strong physics argument in favor of this assumption, except that the effects are largest for these decay channels. 
Our conclusions are mostly the same if one were to pursue the scenario where all allowed two-body decay modes of the heavier neutrinos are present.

Qualitatively, ignoring the relatively much smaller initial flux of $\nu_x$ and $\bar{\nu}_{e,x}$, this is what one expects of the neutronization-burst neutrino-flux if the neutrino lifetime is finite:
\begin{enumerate}
\item[(a)] For the NMO, Dirac neutrinos, and the $\varphi_0$-model, the neutronization-burst neutrinos exit the supernova as $\nu_3$'s. The $\nu_3$ decays into visible and invisible $\nu_1$. While the daughter $\nu_1$ spectrum is softer than the one of the parent, the expected number of events may, in fact, be higher if the $\nu_3$ lifetime is short enough since $|U_{e3}|^2\ll |U_{e1}|^2$. It is possible to confuse this scenario with the IMO-case if the neutrino lifetime is chosen judiciously. 
\item[(b)] For the IMO, Dirac neutrinos, and the $\varphi_0$-model, the neutronization-burst neutrinos exit the supernova as $\nu_2$'s. The $\nu_2$ decays into visible and invisible $\nu_3$. The daughter $\nu_3$ spectrum is softer and $|U_{e3}|^2\ll |U_{e2}|^2$ so one expects a significant suppression of the  neutronization-burst neutrino-flux. It is possible to confuse this scenario with the NMO-case if the neutrino lifetime is chosen judiciously.
\item[(c)] For the NMO, Dirac neutrinos, and the $\varphi_2$-model,  the neutronization-burst neutrinos exit the supernova as $\nu_3$'s. The $\nu_3$ decays into visible and invisible $\bar{\nu}_1$. The $\nu_e$ signal on the Earth is very small, but one anticipates a healthy, softer $\bar{\nu}_e$ signal.
\item[(d)] For the IMO, Dirac neutrinos, and the $\varphi_2$-model, the neutronization-burst neutrinos exit the supernova as $\nu_2$'s. The $\nu_2$ decays into visible and invisible $\bar{\nu}_3$. The $\nu_e$ signal on the Earth is suppressed due to the decay, and one anticipates a small -- $|U_{e3}|^2\ll 1$ -- softer $\bar{\nu}_e$ signal.
\item[(e)] For the NMO, Majorana neutrinos, and the model of interest here (Eq.~(\ref{eq:MajL})), the neutronization-burst neutrinos exit the supernova as $\nu_3$'s. The $\nu_3$ decays into visible $\nu_1$ and $\bar{\nu_1}$ with different, softer spectra. Since $|U_{e3}|^2\ll |U_{e1}|^2$, one may run into an excess of both $\nu_e$ and $\bar{\nu}_e$ at Earth-bound detectors. 
\item[(f)] For the IMO, Majorana neutrinos, and the model of interest here (Eq.~(\ref{eq:MajL})), the neutronization-burst neutrinos exit the supernova as $\nu_2$'s. The $\nu_2$ decays into visible $\nu_3$ and $\bar{\nu_3}$ with different, softer spectra. Since $|U_{e3}|^2\ll |U_{e1}|^2$, one expected a reduced number of $\nu_e$ and $\bar{\nu}_e$ at Earth-bound detectors. 
\end{enumerate} 
Note that, in general, we expect qualitatively different behaviors for the Dirac and Majorana decaying-neutrino scenarios. 

For the original $\nu_x$ spectrum, the impact of neutrino decay is absent for any mass-ordering since, in the regime of interest here, these always exit the supernova as a mixture of the lighter -- assumed to be stable -- mass eigenstates. For anti-neutrinos, the situation is reversed. An antineutrino born as a $\bar{\nu}_e$ during the neutronization burst will exit the supernova as the lightest antineutrino while half of the $\bar{\nu}_x$-born population will exit the supernova as the heaviest antineutrino and can hence subsequentially decay into $\bar{\nu}_1$ ($\bar{\nu}_3$) for the NMO (IMO).

We proceed to discuss more quantitatively the impact of neutrino decay in the measurement of neutrinos from the neutronization burst of the next galactic supernova assuming both the Deep Underground Neutrino Experiment (DUNE) and Hyper-KamiokaNDE (HK) are operational at the time of the momentous event. We first provide the relevant assumptions regarding our detection simulations.

\section{Simulation details}
\label{sec:experiments}

We are interested in the future, very large neutrino detectors DUNE
and HK, expected to come online in the middle of the next decade. The
expected number of supernova neutrino events per unit time $t$ and
unit reconstructed energy $E^r$ at any detector is
\begin{equation}
  \frac{{\rm d}^2N(E^{r},t)}{{\rm dt}\,{\rm d}E^r} = \frac{N_{tg}}{4\pi R^2}\int dE^{t} f_{\nu_{\alpha}}(E^{t},t)\sigma_{\alpha}(E^t)\epsilon(E^{t},E^{r}),
\end{equation}
where $f_{\nu_{\alpha}}$ is the neutrino flux at the Earth,
$\sigma_{\alpha}(E)$ is the relevant detection cross-section, and
$\epsilon(E^{t},E^{r})$ is an energy migration matrix that relates the
true neutrino energy $E^t$ to the reconstructed one. $N_{tg}$ is the
number of targets for the experiment. We have assumed 40~ktons of
liquid argon for DUNE and two water tanks of 187~ktons each for
HK. $R$ is the distance to the supernova. Unless otherwise noted, we
have organized all simulated data into a 2-dimensional array of bins
in energy and time. For the energy, we have considered a bin width of at
least two times the detector energy resolution. For the time evolution
are used 5 bins that account for the first 25~ms of the neutronization
burst.

We have focused only on the dominant processes associated to the
detection of supernova neutrinos at the two next-generation
experiments, described in more detail in what follows. DUNE is a
liquid argon-based experiment \cite{Acciarri:2016crz}, and we only
consider the DUNE far detector given it will to be much larger
than the near detector. For the energies relevant to supernova
neutrinos, DUNE is most sensitive to the $\nu_{e}$ component of the
flux~\cite{Abi:2018dnh}, measured via the charged-current process
\begin{equation}
  \nu_{e} + ^{40}\!\!{\rm Ar} \rightarrow ^{40}\!\!{\rm K}^{*}+e^{-}.
\end{equation}
The neutrino absorption by $^{40}{\rm Ar}$ creates an electron and an
excited nucleus of potassium ($^{40}{\rm K}^{*}$) that will de-excite,
producing a cascade of photons. The overall signal is characterized by
a final state with several low-energy electromagnetic tracks. In order
to properly account for all of these, we make use of
MARLEY~\cite{Gardiner2018}, a Montecarlo event generator that
simulates $\nu_e$ interactions in Argon for energies less than
50~MeV. We use the simulated events to construct the energy migration matrix $\epsilon(E^{t},E^{r})$.

Liquid Argon experiments have proven the capability to observe
electrons and photons in the MeV scale~\cite{Acciarri:2018myr}. For
DUNE, we have assumed a minimum distance of $1.5$~cm traveled by the
electron in order to be detected; such distance translates into an
energy threshold for the electron of $2$~MeV. For photons, the
dominant interaction at those energies is Compton scattering and
photons are observed as isolated blips near the electron track. We
impose the same energy cut in energy as the one for electrons. The
reconstructed neutrino energy is the sum of the reconstructed energies
for all the final state particles, in our case electron and photons
since the remaining recoiling nucleus cannot be observed. The
interaction process proceed with an energy threshold of 4~MeV, which
we consider as threshold for the reconstructed neutrino energy. The
finite energy resolution of the detector introduces an error in the
reconstruction of the neutrino energy. Assuming the same precision as
shown in previous Liquid Argon experiments~\cite{Amoruso:2003sw}, we
have included the energy resolution via Montecarlo integration. The
energy resolution increases as $\sigma_{E} =
0.11\sqrt{E/\text{MeV}} + 0.2 (E/\text{MeV})$, and corresponds to
$5\%$ for $E\sim 10$~MeV. The time resolution for DUNE is expected to
be of the order of $\sim 10$~nsec~\cite{Abi:2018dnh}. Since our
simulated events are organized into time-bins of $5$~ms,
timing-resolution effects are irrelevant.

HK is a water Cherenkov detector. At the MeV scale, the main detection
channel is inverse beta decay (IBD) (Eq.~(\ref{eq:IBD})), so HK is
mainly sensitive to the electron-antineutrino component of the
supernova neutrino flux:
\begin{equation}\label{eq:IBD}
  \overline{\nu}_{e}+p\rightarrow e^{+} + n
\end{equation}
For the cross-section of IBD, we have used the results of the
analytical calculation reported in Ref.~\cite{Strumia:2003zx},
expected to be valid for neutrino energies in the MeV to GeV
range. Finally, we have assumed the energy resolution of HK to be the
same as that of Super-Kamiokande~\cite{Abe:2018uyc}.\footnote{All of
  the results associated to HK also apply to Super-Kamiokande,
  currently taking data, once one takes into account the fact that HK
  is expected to be an order of magnitude larger than
  Super-Kamiokande.} Considering $\sigma_{E} =
0.6\sqrt{E/\text{MeV}}$, for an energy of $10$~MeV, the energy
resolution is of the order of $20\%$. Note that for HK we are assuming
a larger energy resolution which is translated into a larger size of
the energy bins. We have considered a threshold of 3~MeV in the energy
measured, which is imposed by the detector capabilities to observe
neutrinos~\cite{Abe:2018uyc}. To correlate the neutrino energy with
the energy reconstructed by the detector, we made a Montecarlo
integration assuming a Gaussian distribution of the energies measured
by the detector, where every event is weighted by the differential
cross-section~\cite{Strumia:2003zx}. The time resolution for HK is of
order $ 5$~nsec~\cite{Abe:2018uyc}, negligible compared to our 5~ms
time-bins.

In addition to IBD, the neutrino--electron scattering channel also contributes to the detection of SN neutrinos in HK.  Given the large size of the detector, one can expect $\sim$ 23 (55) events for NMO (IMO) in one tank, for a SN occuring at 10 kpc. This channel can, in principle, make it possible to observe the neutronization peak, if proper background subtraction can be made.  However, the neutrino--electron scattering channel is sensitive to neutrinos of all flavors. 
Due to the difficulties in disentangling the $\nu_{e} e$ scattering from the other neutrino flavors as well as the IBD events, we consider only the IBD channel for HK. The identification of the
background via the neutron tagging by adding Gadolinium requires a more dedicated analysis, something that we will consider in future extensions of this work.

We make use of a $\chi^2$ analysis in order to compare different
hypotheses and address different physics questions. We assume a
  Gaussian distribution for the $\chi^2$. For concreteness, the
  $\Delta\chi^2$ for a set of values of the parameters gives us the
  significance over the test hypothesis. We
bin our simulated events in energy and time, as discussed above, and
marginalized over the different nuisance parameters in order to
account for different systematic and statistical effects. For the
remainder of this manuscript, unless otherwise noted, we assume that
the overall normalization of the supernova neutrino flux is known at
the 40\% level (one sigma). For the neutrino mixing parameters,
  we use the results of the global fit reported by the NuFit collaboration
\cite{Esteban:2018azc}, and marginalize over the reported
uncertainties for the relevant mixing parameters --
$\theta_{12}$ and $\theta_{13}$.

\section{Impact of neutrino decay at DUNE and Hyper-Kamiokande}
\label{sec:results}
In this section, we explore some of the consequences of the neutrino decay hypothesis to future data from DUNE and HK.  For concreteness, we concentrate on the hypothesis that the neutrinos are Dirac fermions and only the heaviest neutrino decays to the lightest neutrino. Therefore, in the NMO (IMO), the $\nu_3\,(\nu_2)$ decays to $\nu_1\,(\nu_3)$, similarly for antineutrinos. We will also concentrate on the $\varphi_0$-model with $g_{ij}=g_{ji}$. Since we are assuming that DUNE is only sensitive to the $\nu_e$-component of the neutronization-burst neutrinos on Earth and HK is only sensitive to the $\bar{\nu}_e$-component, DUNE is expected to play a more significant role. It is clear from earlier discussions that the roles of the two detectors would be reversed in the $\varphi_2$-model. Certain aspects of the hypothesis where the neutrinos are Majorana fermions -- Eq.~(\ref{eq:MajL}) -- were explored earlier in the literature \cite{Ando:2004qe}. We return to the Majorana fermion versus Dirac fermion discussion later.

\begin{figure}[!t]
\includegraphics[width=0.48\textwidth]{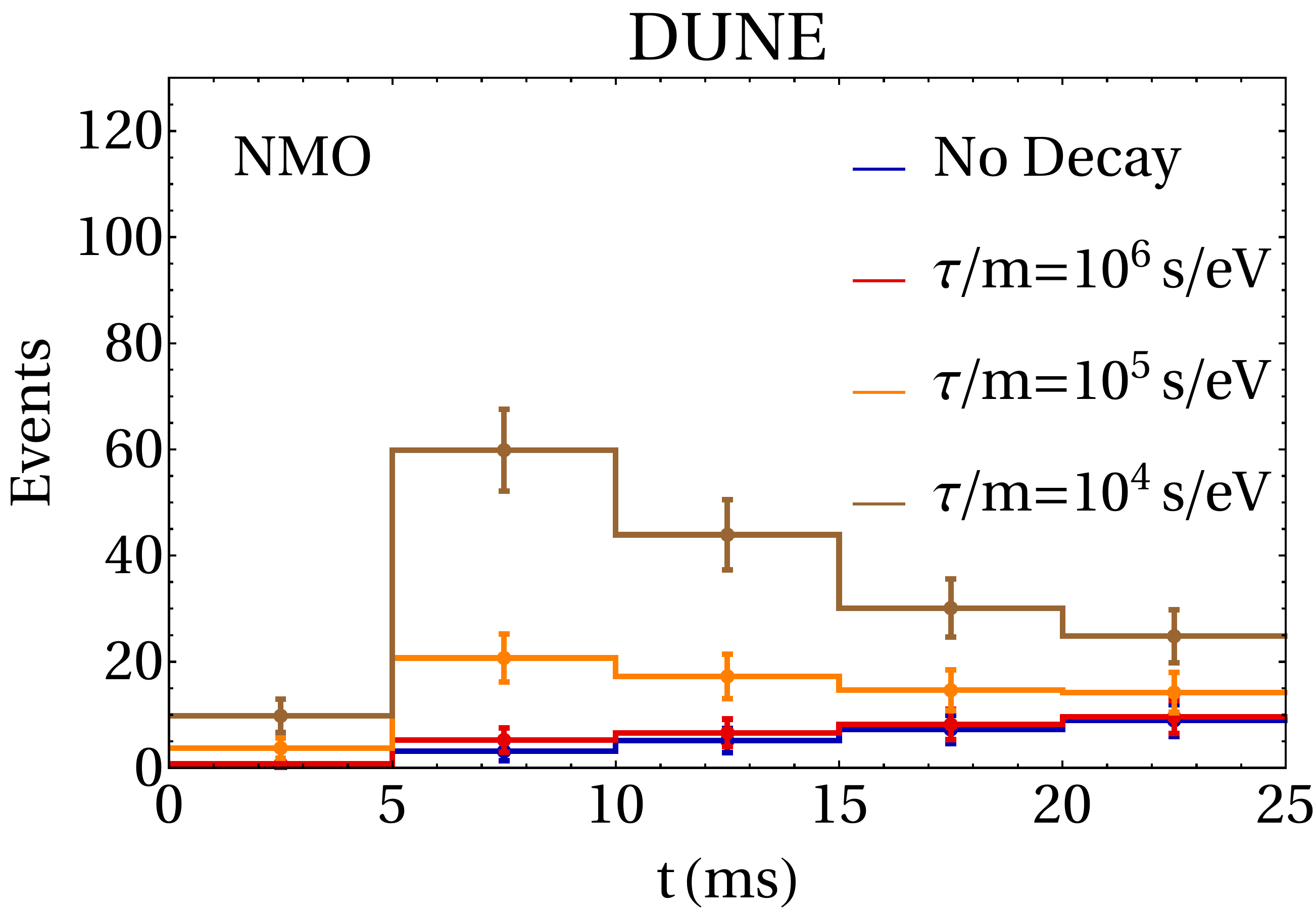}\,\quad\includegraphics[width=0.48\textwidth, height=0.345\textwidth]{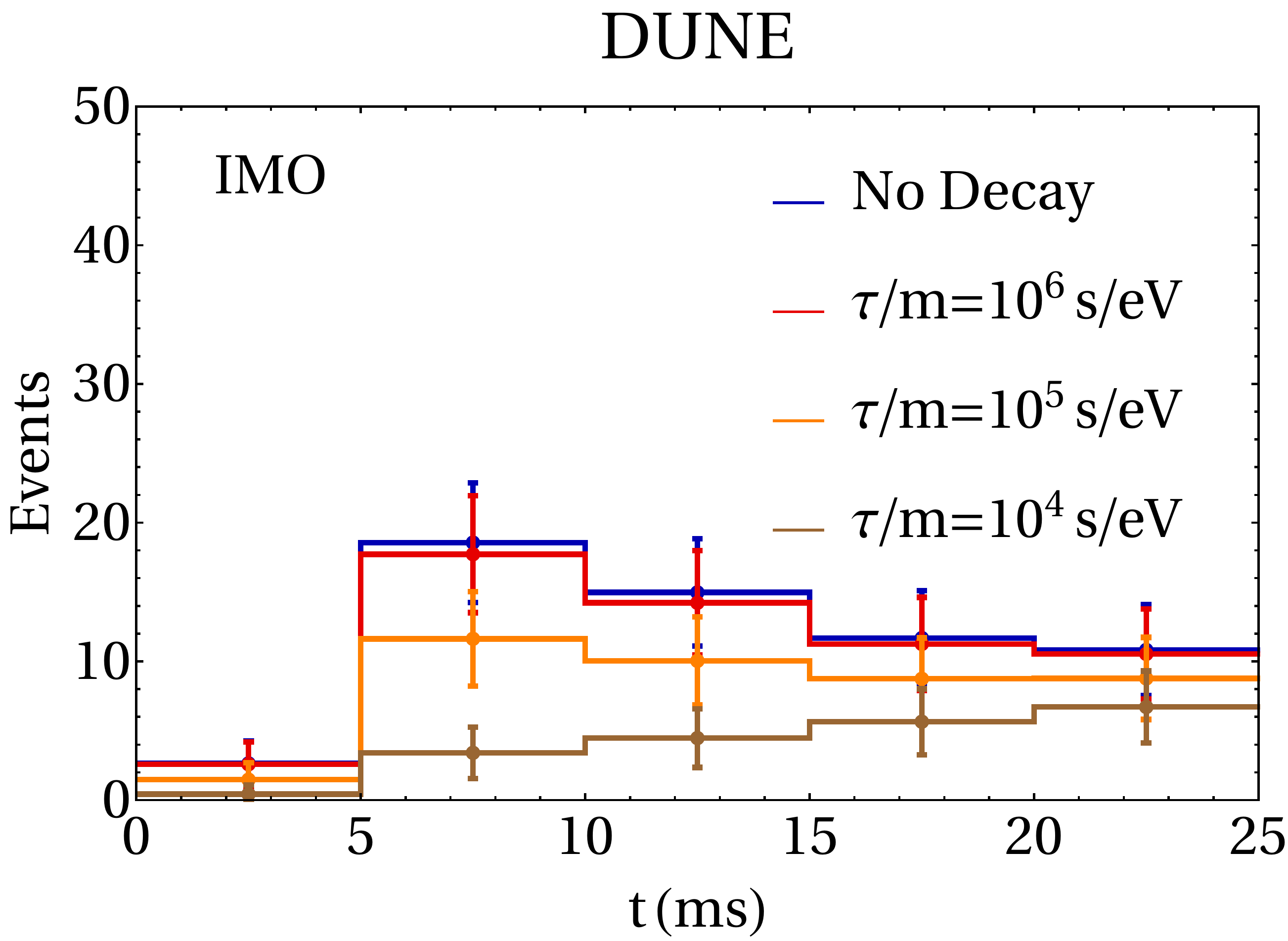}
\caption{Expected event rates as a function of time in DUNE (40~ktons of liquid argon), for different values of the lifetime of the heaviest neutrino mass eigenstate, for the NMO (left) and the IMO (right), for a supernova explosion 10~kpc away. We include all events that deposit at least 4~MeV of reconstructed energy in the detector. The values of the neutrino oscillation parameters are the best-fit ones tabulated in \cite{Esteban:2018azc}.  The neutrinos are assumed to be Dirac fermions and the neutrino decay is described by Eq.~(\ref{eq:DiracL}).}
\label{fig:Decay_time}
\end{figure}

Assuming a supernova explosion 10~kpc away, Fig.\,\ref{fig:Decay_time} depicts the expected number of events of all reconstructed neutrino energies at DUNE, divided into bins of 5~ms, for both neutrino mass orderings and for different values of the lifetime-to-mass ratio $\tau/m$ of the heaviest neutrino. For the IMO (right panel of Fig.\,\ref{fig:Decay_time}), the neutronization-burst  peak at 10~ms after bounce is visible for very long-lived neutrinos, while there is no peak for the NMO (left panel of Fig.\,\ref{fig:Decay_time}). For shorter neutrino lifetimes, for the NMO, a peak develops and becomes quite pronounced due to the presence of the easier-to-detect $\nu_1$ daughters. The situation is reversed for the IMO: for shorter neutrino lifetimes, the neutrino decay erases the neutronization-burst  peak. 
\begin{figure}[!t]
\includegraphics[width=0.48\textwidth]{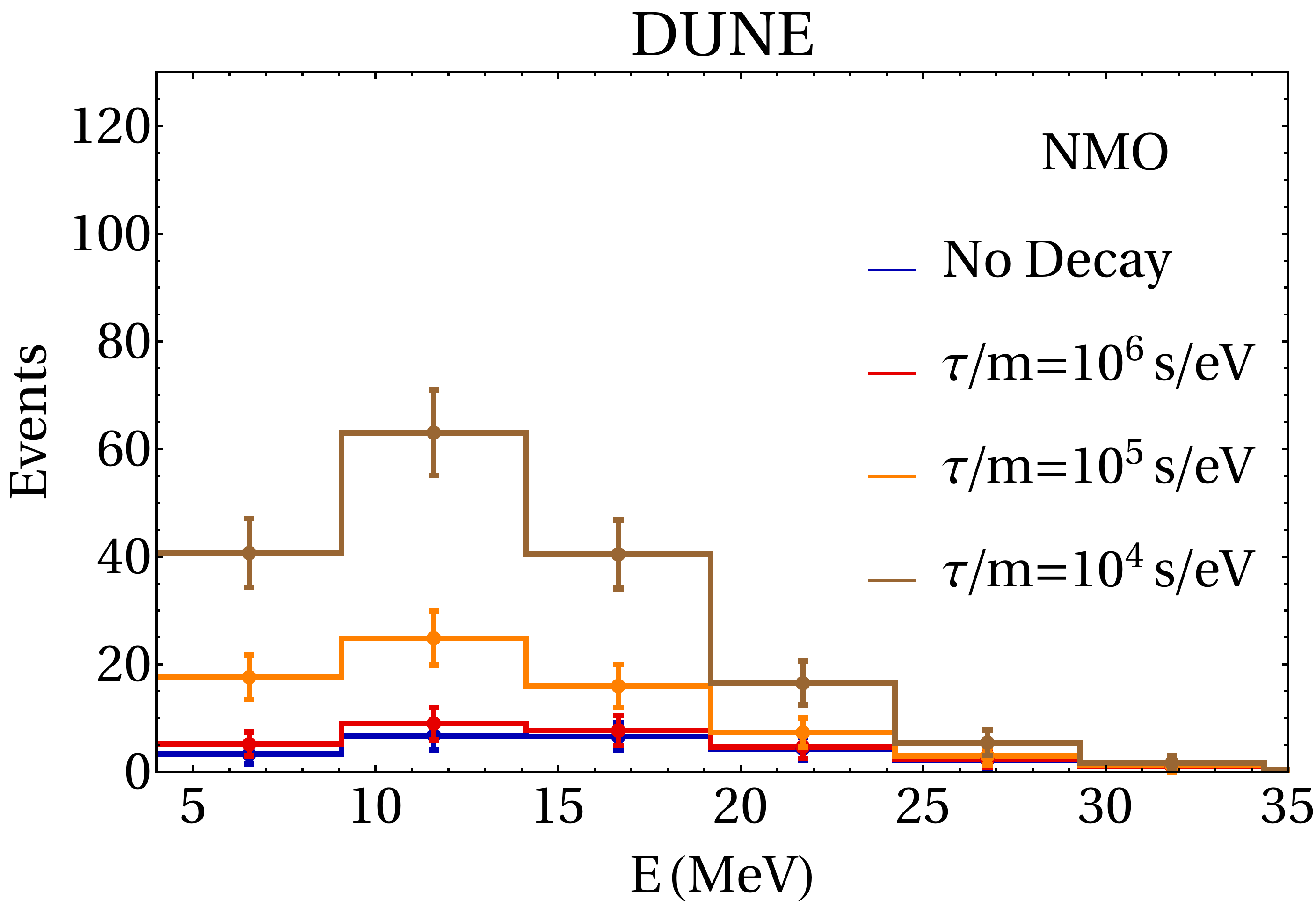}\,\quad\includegraphics[width=0.48\textwidth,  height=0.344\textwidth]{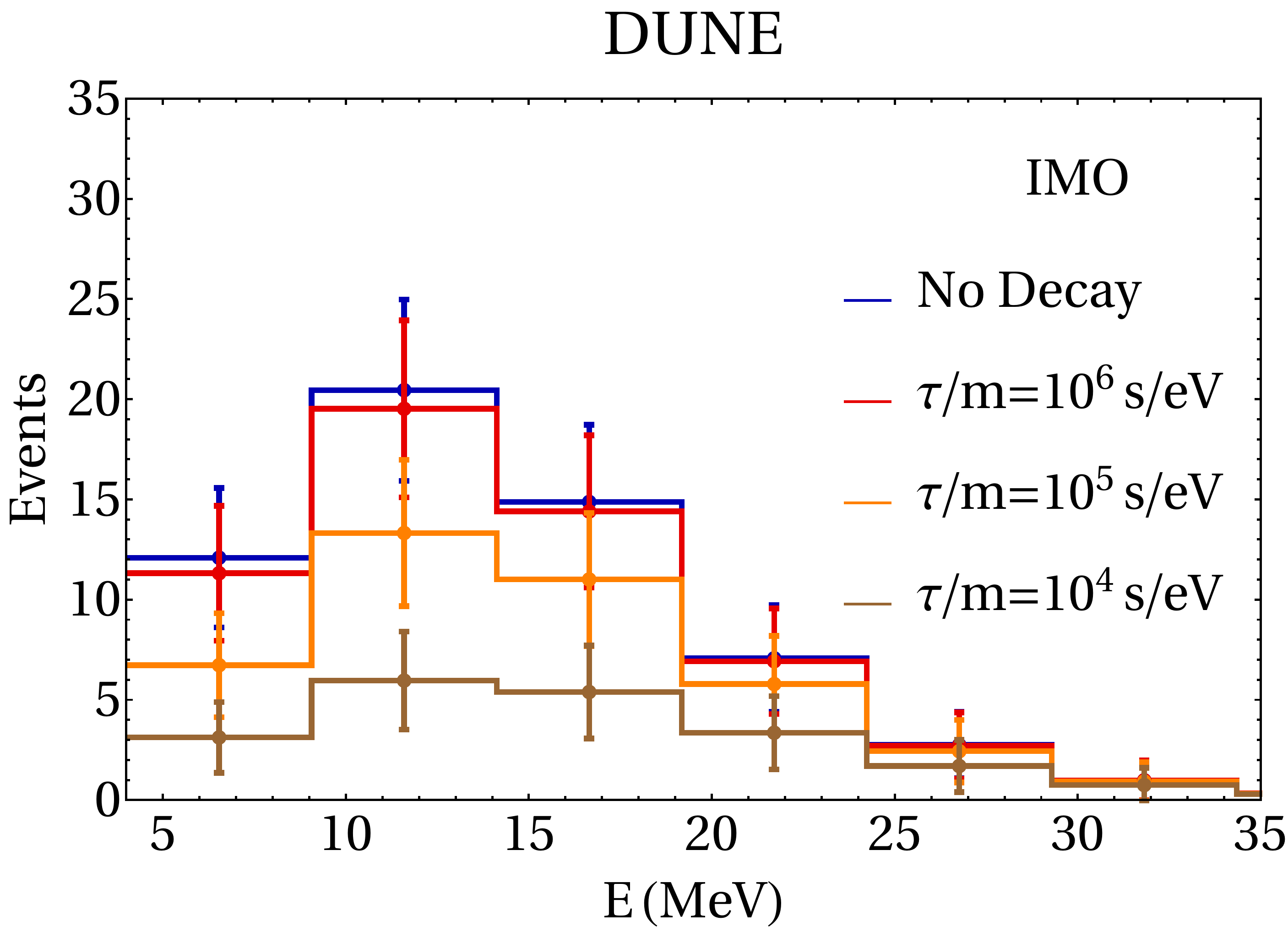}
\caption{Expected event rates as a function of energy in DUNE (40~ktons of liquid argon), for different values of the lifetime of the heaviest neutrino mass eigenstate, for the NMO (left) and the IMO (right), for a supernova explosion 10~kpc away. We include all events that arrive during the neutronization-burst (first 25~ms). The values of the neutrino oscillation parameters are the best-fit ones tabulated in \cite{Esteban:2018azc}.  The neutrinos are assumed to be Dirac fermions and the neutrino decay is described by Eq.~(\ref{eq:DiracL}).}
\label{fig:Decay_En}
\end{figure}

Assuming a supernova explosion 10~kpc away, Fig.~\ref{fig:Decay_En} depicts the number of events observed inside the first 25~ms time-window in bins of reconstructed neutrino energy for the NMO (left panel), and the IMO (right panel). The same pattern described in the previous paragraph is observed. The $\nu_e$ spectrum for the IMO is suppressed as the decay rate increases, while that of the NMO is enhanced. The neutrino energy spectrum softens slightly as the lifetime decreases. The effect is not very pronounced because the visible daughter-neutrino energy distribution (linearly) peaks at the parent energy, see Eq.~(\ref{eq:energydist}) ($\langle E_{\rm daughter} \rangle=(2/3) E_{\rm parent}$).

\subsection{Normal versus Inverted Mass Ordering}

\begin{figure}[!t]
\includegraphics[width=0.85\textwidth]{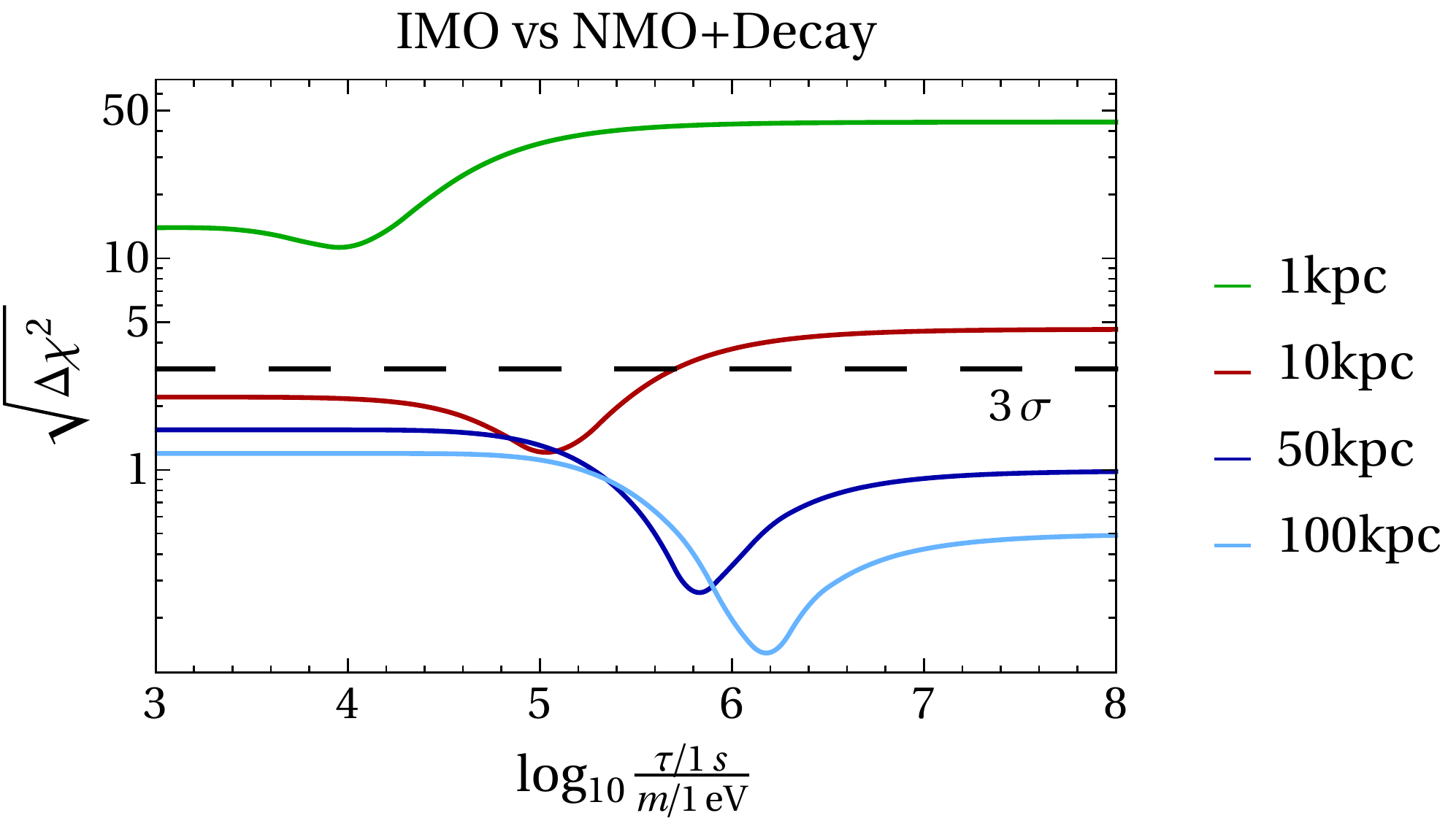}
\caption{$\sqrt{\Delta\chi^2}$ as a function of the hypothetical $\nu_3$ lifetime-over-mass, obtained when comparing simulated DUNE data consistent with the IMO and effectively stable neutrinos with the hypothesis that the neutrino mass ordering is normal and the $\nu_3$ lifetime is finite, for different distances between the Earth and the supernova explosion. Data are binned in both time and energy, as described in the text. We assume a 40\% uncertainty in the overall neutrino flux, and marginalize over the neutrino oscillation parameters (except for the mass ordering).  The values of the neutrino oscillation parameters -- best-fit values and uncertainties -- are tabulated in \cite{Esteban:2018azc}. The neutrinos are assumed to be Dirac fermions and the neutrino decay is described by Eq.~(\ref{eq:DiracL}).}
\label{fig:ChiSq}
\end{figure}

The past discussion reveals that, qualitatively, a normal neutrino mass ordering can be mimicked by an inverted one if one allows the neutrinos to decay, and vice-versa. We next address this issue more quantitatively by simulating DUNE data consistent with the IMO and attempting to fit the data with a NMO while allowing the heavier neutrino to decay, as described above. The results of this exercise are depicted in Fig.\,\ref{fig:ChiSq}, which displays $\sqrt{\Delta \chi^2}$ as a function of the $\nu_3$ lifetime-over-mass for different supernova--Earth distances. For explosions that are very far away -- over 50~kpc -- the two mass-orderings are indistinguishable at the two-sigma level regardless of the $\nu_3$ lifetime. This is related to the simple fact that there is not enough statistics to distinguish one mass-ordering from the other at DUNE. For explosions that are very nearby -- say, 1~kpc -- it is not possible to mimic the IMO with the NMO plus a finite lifetime for $\nu_3$. This is a reflection of the fact that the energy spectrum (and, to a lesser extent, time) is distorted enough by the neutrino decay that the two hypothesis are distinguishable given enough statistics. At intermediate distances -- e.g., 10~kpc -- it is possible to mimic an IMO with a NMO plus a finite lifetime of $\nu_3$ if the lifetime is chose judiciously.  Fig.\,\ref{fig:ChiSq} reveals that the five-sigma distinction between the IMO and the NMO is reduced to under two-sigma if, in the NMO, the $\nu_3$ is allowed to decay into a $\nu_1$ plus a massless scalar with $\tau_3/m_3\sim 10^{5}$~s/eV. The shape of the different curves in Fig.\,\ref{fig:ChiSq} is easy to understand. The value of $\tau_3/m_3$ for which decay effects are most interesting depends on the distance to the supernova. The positions of the minima in Fig.\,\ref{fig:ChiSq} follow Eq.~(\ref{eq:estimate}).

\subsection{Constraining the $\nu_h$ Lifetime}

\begin{figure}[!t]
\includegraphics[width=0.85\textwidth]{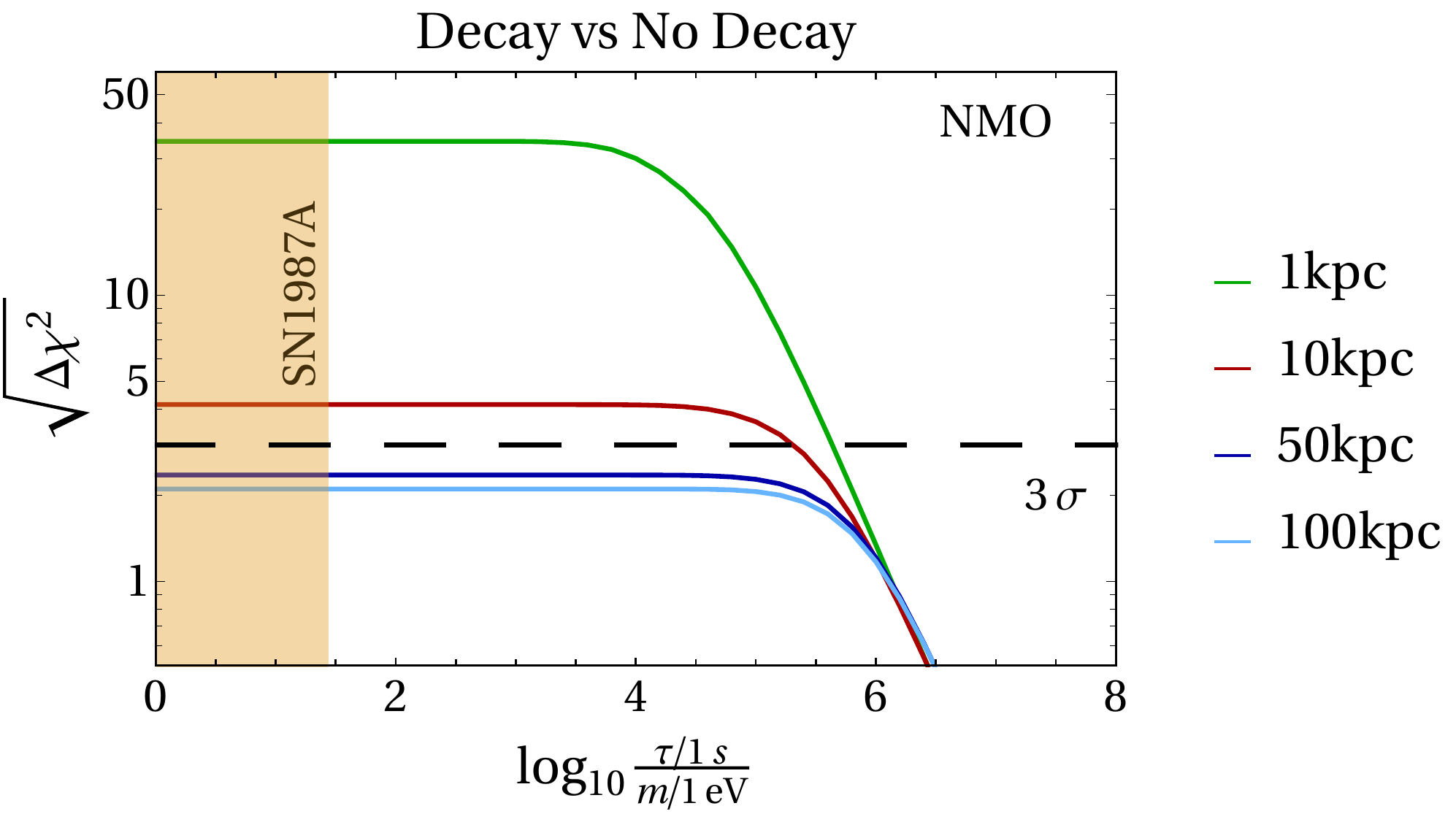}
\caption{$\sqrt{\Delta\chi^2}$ as a function of the hypothetical $\nu_3$ lifetime-over-mass, obtained when comparing simulated DUNE data consistent with effectively stable neutrinos with the hypothesis that the $\nu_3$ lifetime is finite, for different distances between the Earth and the supernova explosion. The neutrino mass ordering is assumed to be normal and known. Data are binned in both time and energy, as described in the text. We assume a 40\% uncertainty in the overall neutrino flux, and marginalize over the neutrino oscillation parameters (except for the mass ordering).  The values of the neutrino oscillation parameters -- best-fit values and uncertainties -- are tabulated in \cite{Esteban:2018azc}. The neutrinos are assumed to be Dirac fermions and the neutrino decay is described by Eq.~(\ref{eq:DiracL}). The colored shaded region is excluded by supernova cooling bounds using data from SN1987A observations \cite{Kachelriess:2000qc,Farzan:2002wx}.}
\label{fig:ChiSqTau}
\end{figure}

When the neutrinos from the next galactic supernova arrive at DUNE and HK, it is possible that the neutrino mass ordering will be known. In this case, one can use the supernova neutrino data to constrain the neutrino lifetime. Assuming the mass ordering is known to be normal, currently favored at the three sigma level by the neutrino oscillation data \cite{Esteban:2018azc}, Fig.~\ref{fig:ChiSqTau} depicts $\sqrt{\Delta \chi^2}$ as a function of $\tau_3/m_3$ assuming DUNE observed events from a supernova explosion at different distances and that the neutrino lifetime is infinitely long. For supernova explosions very far away, DUNE cannot distinguish a very long-lived $\nu_3$ from a very short-lived one due to the lack of statistics. For near-enough explosions, DUNE can rule out, at the three-sigma level, lifetimes shorter than $\sim 3\times 10^{5}$~s/eV. This sensitivity is vastly superior to current bounds from long-baseline and solar neutrino data, as anticipated. 

It is easy to understand the behavior of the various curves in Fig.~\ref{fig:ChiSqTau}. For small enough values of $\tau/m$, $\sqrt{\Delta \chi^2}$ does not depend on it because the decays are too fast. On the other hand, for large values of $\tau/m$, we expect relative deviations of the number of expected neutronization-burst neutrinos, relative to infinitely long lifetimes, to vary proportional to $Lm/E\tau$. This behavior is readily observed for different values of $L$ in Fig.~\ref{fig:ChiSqTau}, keeping in mind that the energies of interest are approximately the same. The $L$-dependent change in behavior follows Eq.~(\ref{eq:estimate}). For large $\tau/m$, the sensitivity -- very small -- is mostly independent on $L$. The reason for this is as follows. The total number of events is proportional to $(1/L)^2$ while, as highlighted above, the relative change in the number of events is proportional to $L$. Hence, the absolute change in the number of events -- decay hypothesis versus stable hypothesis -- decreases like $1/L$. On the other hand, the statistical uncertainty of the ``data'' decreases like $\sqrt{(1/L^2)}$. The statistical power to discriminate one hypothesis from the other is proportional to $(1/L)/(1/L)$ and hence $L$-independent. 
 
Fig.~\ref{fig:ChiSqTau} reveals that if a supernova explodes 10~kpc away, DUNE data will be able to clearly distinguish short-lived $\nu_3$ ($\tau_3/m_3\ll 10^{5}$~s/eV) from a long-lived one ($\tau_3/m_3\gg10^{5}$~s/eV). If the $\nu_3$ lifetime happens to be $\tau_3/m_3\sim 10^{5}$~s/eV, the energy and time distribution of the neutrinos from the neutronization burst are capable of providing more detailed information about $\tau_3/m_3$. The solid, blue line in Fig.~\ref{fig:ChiSqDvsM} depicts $\sqrt{\Delta \chi^2}$ as a function of $\tau_3/m_3$, obtained by analyzing data at DUNE consistent with $\tau_3/m=10^{5}$~s/eV (explosion 10~kpc away), assuming the mass ordering is known to be normal. At the two-sigma level, DUNE can establish that $\tau_3/m_3$ is neither very short nor very long and it can measure $\tau_3/m_3=(1.0^{+0.6}_{-0.4})\times 10^{5}$~s/eV (one sigma).

\subsection{Dirac versus Majorana: DUNE $\&$ Hyper-Kamiokande}

If the neutrinos decay, the neutronization burst may also prove to be an excellent laboratory for testing the nature -- Majorana versus Dirac -- of the neutrino. As discussed in Sec~\ref{sec:neutrinodecay}, the decays of Majorana and Dirac neutrinos are qualitatively different if the daughter neutrino can be subsequently measured. For example, if the neutrinos are Dirac fermions and neutrino decay is governed by Eq.\,(\ref{eq:DiracL}) ($\varphi_0$ scenario) neutrinos decay into visible or invisible neutrinos while antineutrinos decay into visible or invisible antineutrinos. On the other hand, Majorana neutrinos, assuming the decay is governed by Eq.\,(\ref{eq:MajL}), if produced in charged-current processes absorbing negatively-charged leptons, will decay, half of the time into ``neutrinos'' and half of the time into ``antineutrinos.'' In this case, all daughter fermions are visible.

\begin{figure}[!t]
\includegraphics[width=0.48\textwidth]{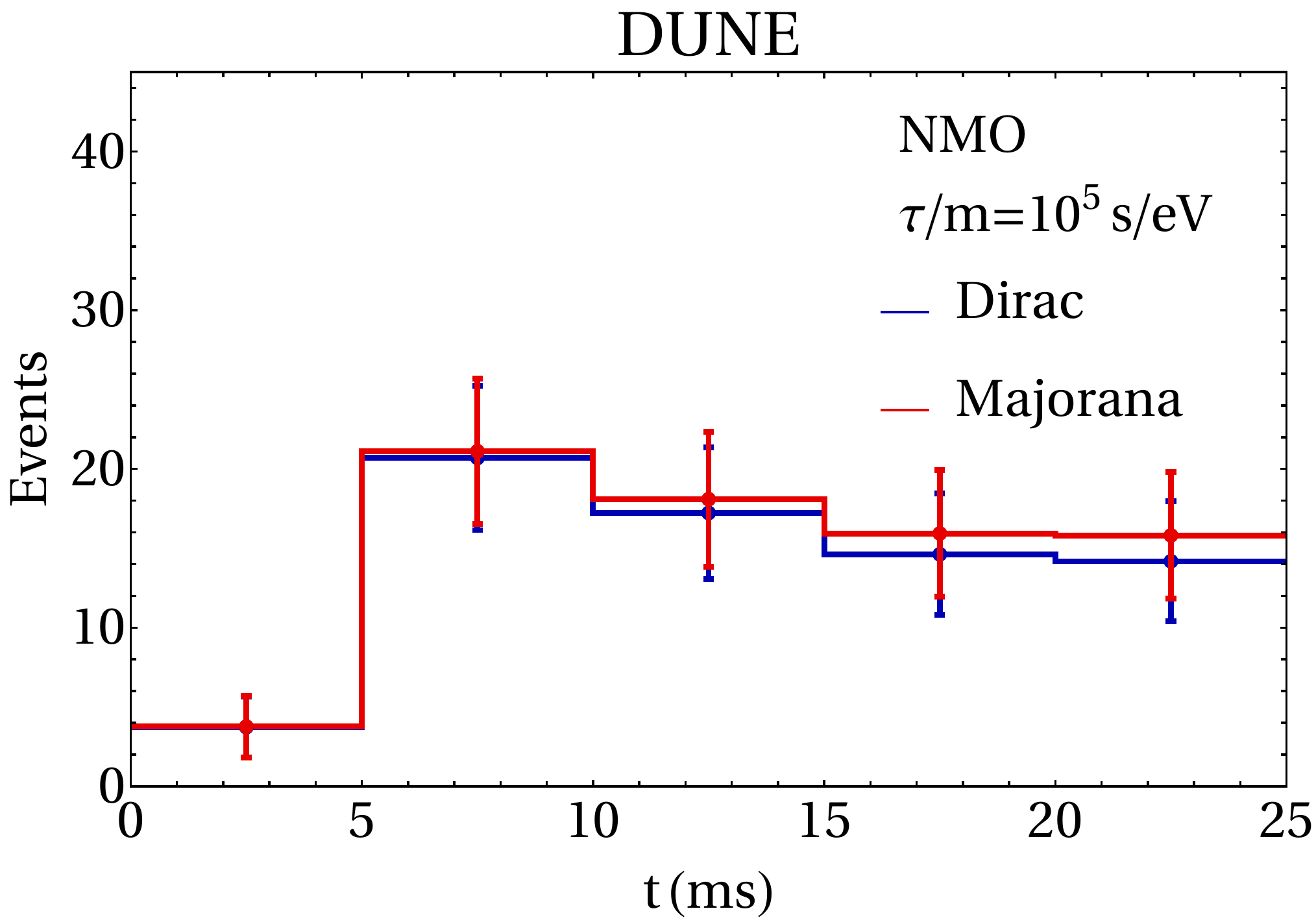}\,\quad\includegraphics[width=0.48\textwidth, height=0.334\textwidth]{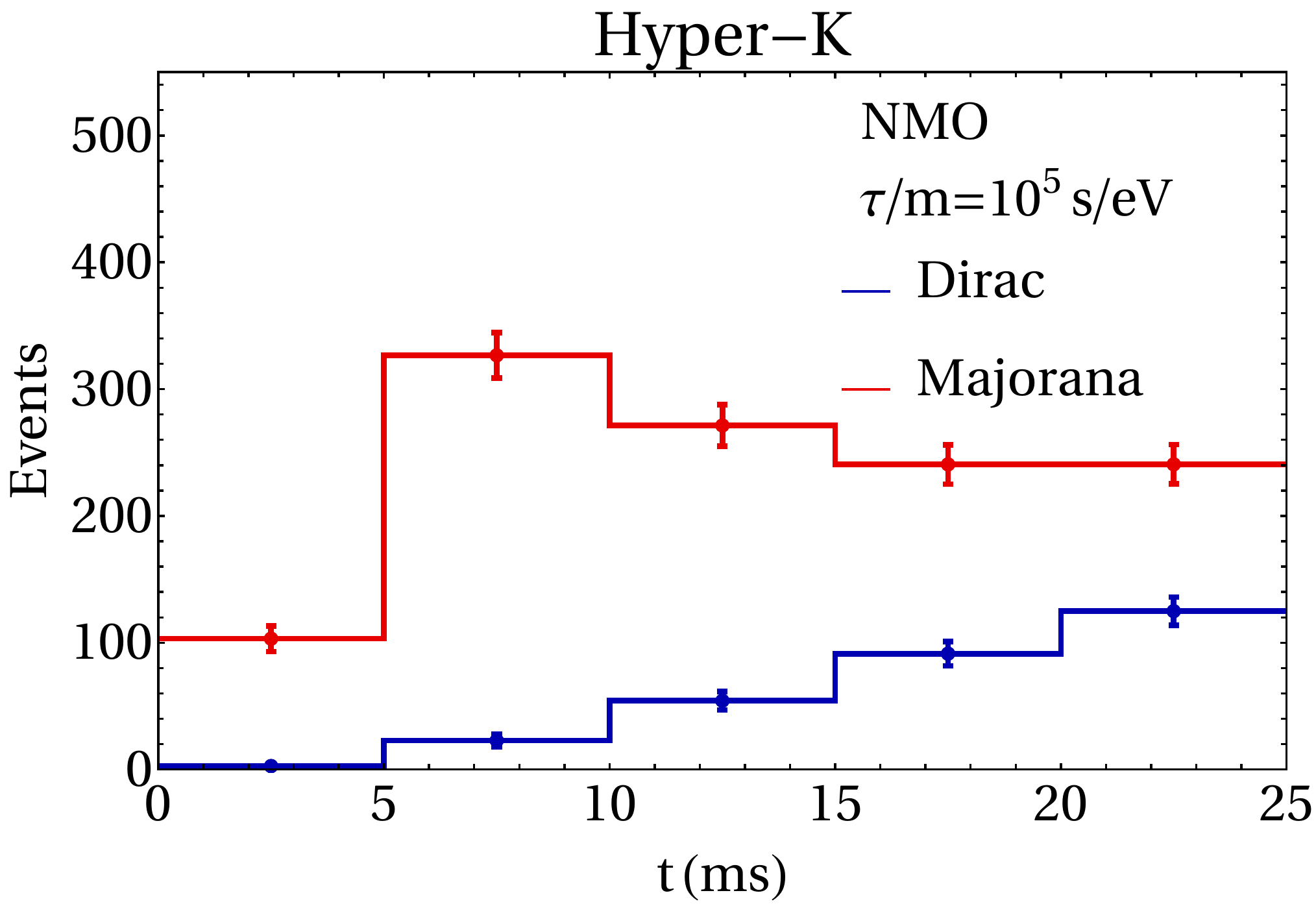}
\caption{Expected event rates from a supernova explosion 10~kpc away as a function of time in DUNE (left, 40~ktons of liquid argon) and HK (right, 374~ktons of water), for decaying Dirac (blue) and Majorana $\nu_3$ (red), for $\tau_3/m_3=10^5$~s/eV. We include all events that deposit at least 4~MeV (3~MeV) of reconstructed energy in DUNE (HK). The values of the neutrino oscillation parameters are the best-fit ones tabulated in \cite{Esteban:2018azc} and the mass ordering is normal.  Dirac neutrino decay is described by Eq.~(\ref{eq:DiracL}) while Majorana neutrino decay is described by Eq.~(\ref{eq:MajL}).}
\label{fig:DUNEvsSK_time}
\end{figure}

\begin{figure}[!h]
\includegraphics[width=0.48\textwidth]{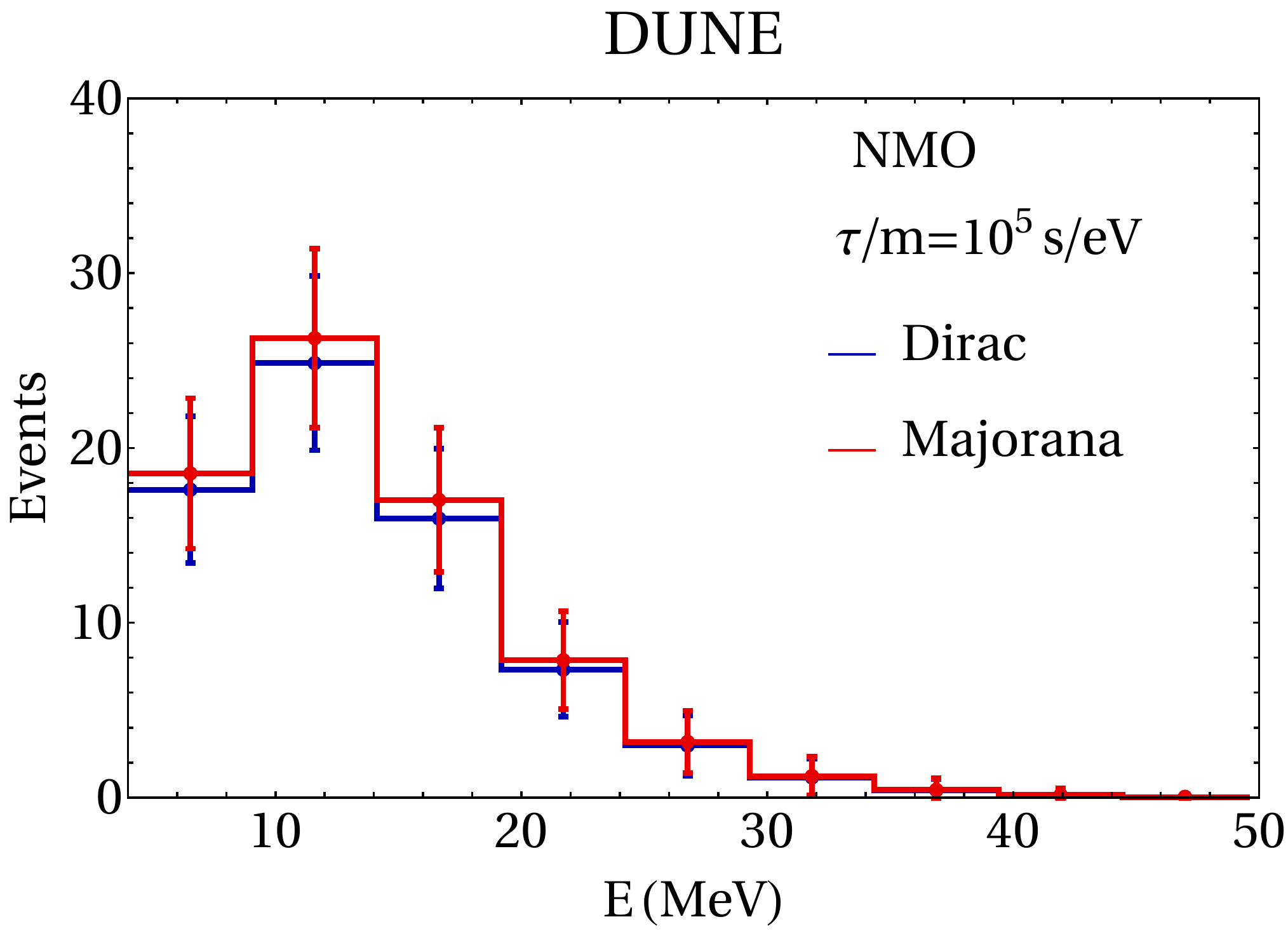}\,\quad\includegraphics[width=0.48\textwidth, height=0.346\textwidth]{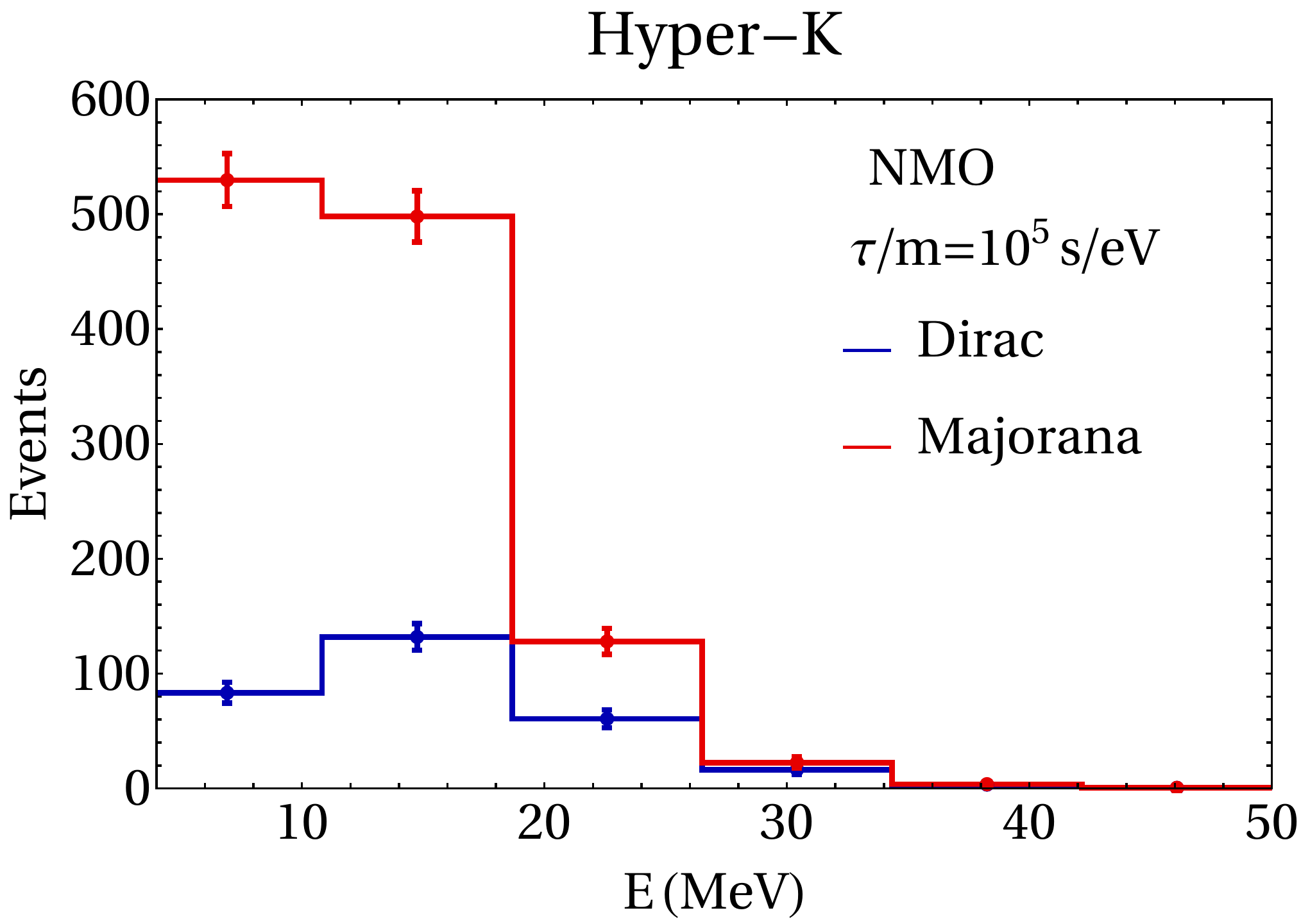}
\caption{Expected event rates from a supernova explosion 10~kpc away as a function of energy in DUNE (left, 40~ktons of liquid argon) and HK (right, 374~ktons of water), for decaying Dirac (blue) and Majorana $\nu_3$ (red), for $\tau_3/m_3=10^5$~s/eV. We include all events that arrive during the neutronization-burst (first 25~ms). The horizontal scales do not start at $E=0$, reflecting the energy thresholds for the experiments. The values of the neutrino oscillation parameters are the best-fit ones tabulated in \cite{Esteban:2018azc} and the mass ordering is normal.  Dirac neutrino decay is described by Eq.~(\ref{eq:DiracL}) while Majorana neutrino decay is described by Eq.~(\ref{eq:MajL}).}
\label{fig:DUNEvsSK_En}
\end{figure}

The qualitative consequences of decaying neutrinos from the neutronization-burst of a supernova explosion were spelled out in Sec.~\ref{sec:decaysupernova}. Since we are assuming that DUNE is, at leading order, sensitive to the $\nu_e$ population on the Earth while HK is sensitive to the $\bar{\nu}_e$ population, the neutrino decay hypothesis will impact only DUNE data if the neutrinos are Dirac fermions, while expectations at HK are expected to be qualitative different if the decaying neutrinos are Majorana fermions. This is illustrated in Figures~\ref{fig:DUNEvsSK_time} and \ref{fig:DUNEvsSK_En}.

Fig.\,\ref{fig:DUNEvsSK_time} depicts the expected number of events of all neutrino energies at DUNE (left) and HK (right), for a supernova exploding 10~kpc away, the NMO, and $\tau_3/m_3=10^{5}$~s/eV, assuming Dirac neutrinos (and Eq.\,(\ref{eq:DiracL}) together with $g_{ij}=g_{ji}$) or Majorana neutrinos (and Eq.\,(\ref{eq:MajL})). The data are virtually the same for the two hypothesis at DUNE,\footnote{The slight excess of $\nu_e$ events at DUNE is due to the decay of the ``antineutrinos:'' $\bar{\nu}_3\to\nu_1+\varphi$, keeping in mind that we are assuming $g_{13}=g_{31}$ in the Dirac $\varphi_0$-model, so half of the daughters of the Dirac $\nu_3$ decay have the ``wrong'' helicity and are therefore invisible.} but the situation is qualitatively different at HK. If the neutrinos are Majorana fermions, half of the time, the $\nu_3\to\bar{\nu}_1+\varphi$. The daughter $\bar{\nu}_1$ behave, $|U_{e1}|^2$ of the time, as $\bar{\nu}_e$ and hence lead to a large signal at HK. 

The same behavior is observed in Fig.\,\ref{fig:DUNEvsSK_En}, which depicts the energy spectrum of the neutrinos arriving in the first 25~ms at DUNE (left) and HK (right), for a supernova exploding 10~kpc away, the NMO, and $\tau_3/m_3=10^{5}$~s/eV, assuming Dirac neutrinos (and Eq.\,(\ref{eq:DiracL})) or Majorana neutrinos (and Eq.\,(\ref{eq:MajL})). While the ``wrong-helicity'' (``antineutrino'') population has a markedly softer energy spectrum, as discussed earlier, it still leads to a significant excess of events in HK.

\begin{figure}[!t]
\includegraphics[width=1.\textwidth]{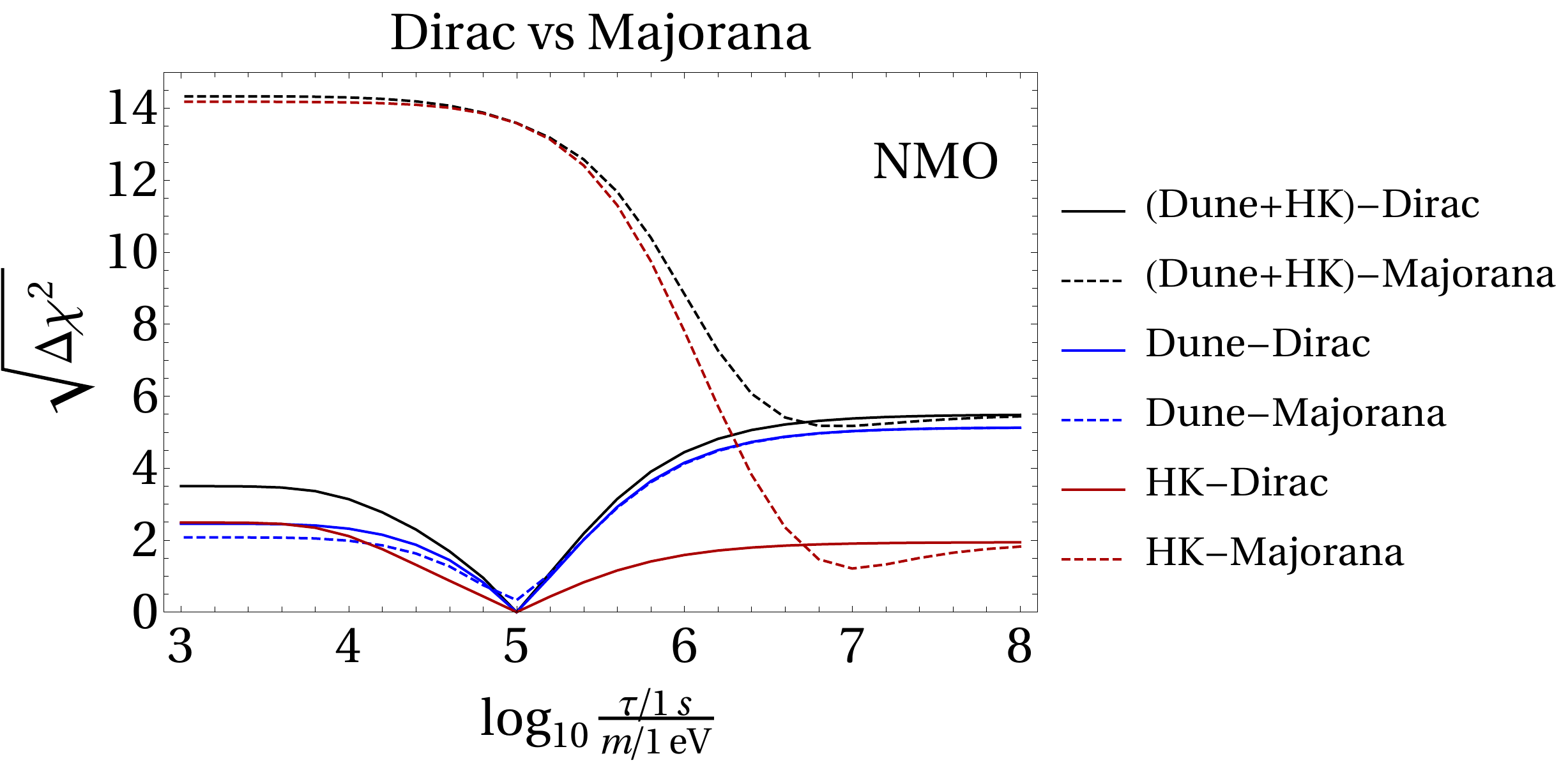}
\caption{$\sqrt{\Delta\chi^2}$ as a function of the hypothetical $\nu_3$ lifetime-over-mass, obtained when comparing simulated DUNE or HK data consistent with a decaying Dirac $\nu_3$ from a supernova explosion 10~kpc away, for $\tau_3/m_3=10^5$~s/eV, with different hypothesis regarding the $\nu_3$ lifetime and the nature of the neutrinos. The neutrino mass ordering is assumed to be normal and known. Data are binned in both time and energy, as described in the text. We assume a 40\% uncertainty in the overall neutrino flux -- independent at DUNE and HK --  and marginalize over the neutrino oscillation parameters (except for the mass ordering).  The values of the neutrino oscillation parameters -- best-fit values and uncertainties -- are tabulated in \cite{Esteban:2018azc}. Dirac neutrino decay is described by Eq.~(\ref{eq:DiracL}) while Majorana neutrino decay is described by Eq.~(\ref{eq:MajL}).}
\label{fig:ChiSqDvsM}
\end{figure}

We simulated data at DUNE and HK consistent with a supernova explosion 10~kpc away assuming the neutrino mass ordering is known to be normal and that $\nu_3$ is a decaying Dirac fermion with $\tau_3/m_3=10^{5}$~s/eV. We analyze subsets of the data under different hypotheses regarding the nature of the neutrinos -- decaying Majorana or Dirac $\nu_3$ -- and display our results in Fig.~\ref{fig:ChiSqDvsM}. As hinted in Figs.~\ref{fig:DUNEvsSK_time} and \ref{fig:DUNEvsSK_En}, DUNE data alone (blue lines) cannot distinguish Dirac (solid line) from Majorana (dashed line) neutrinos. HK data alone (red lines), on the other hand, does a slightly better job but can only distinguish a decaying Dirac (solid line) $\nu_3$ from a very long-lived Majorana (dashed line) one at the one sigma level. The solid red line reveals that HK does not have enough sensitivity to distinguish a decaying Dirac neutrino from a stable one. It does, however, have the same capability to distinguish a fast-decaying Dirac $\nu_3$ from one with $\tau_3/m_3=10^{5}$~s/eV as DUNE. The reason is that, in spite of the fact that HK does not see the dominant neutrinos from the deleptonization process, it is large enough to see a very healthy sample of antineutrinos born as $\bar{\nu}_e$ and $\bar{\nu}_x$, even if one restricts the analysis only to the neutronization-burst period (see Figs.~\ref{fig:DUNEvsSK_time} and \ref{fig:DUNEvsSK_En}.). 

The combination of DUNE and HK data leads to sensitivity that dwarfs that of each individual data set; the two detectors complement one another very well. The results of combined DUNE and HK data are depicted in Fig.~\ref{fig:ChiSqDvsM} (black lines). Since the production mechanism responsible for the $\nu_e$ population and that of $\nu_x, \bar{\nu}_e, \bar{\nu}_x$ are qualitatively different during the neutronization-burst -- $\nu_e$
come predominantly from $e^-+p\to \nu_e+n$ while the other species come from so-called thermal processes -- we marginalize over independent normalization uncertainties -- 40\% each -- for DUNE and HK in order to be as conservative as possible. If we assign identical normalization factors to the two data sets, the sensitivity is enhanced by a couple of units of $\sqrt{\Delta \chi^2}$. In this case, combined data from DUNE and HK can distinguish the decaying Dirac $\nu_3$ from a Majorana $\nu_3$ at the five sigma level. Furthermore, combined DUNE and HK data can distinguish a fast-decaying $\nu_3$ from one with $\tau_3/m_3=10^{5}$~s/eV at around the four sigma level. 

In the case of HK and the Majorana neutrino hypothesis, Fig.~\ref{fig:ChiSqDvsM} reveals a shallow minimum for $\tau_3/m_3\sim 10^7$~s/eV. This is qualitatively easy to understand. For smaller values of $\tau_3/m_3$, the expected number of events at HK is way too large and hence the disagreement with the simulated data -- Dirac neutrinos and $\tau_3/m_3\sim 10^5$~s/eV -- is strong, as depicted in Figs.~\ref{fig:DUNEvsSK_time} and \ref{fig:DUNEvsSK_En}. The data are also modestly inconsistent with a stable $\nu_3$, as one can see clearly in Fig.~\ref{fig:ChiSqDvsM}. The reason is that, for Dirac neutrinos and $\tau_3/m_3\sim 10^5$~s/eV, the expected number of events at HK is larger than what one get when $\nu_3$ is stable. When $\tau_3/m_3\sim 10^7$~s/eV and the neutrinos are Majorana fermions, a tiny fraction of $\nu_3\to\bar{\nu}_1+\varphi$. Since the primordial $\nu_e$ flux is much larger than that of $\nu_x,\bar{\nu}_e,\bar{\nu}_x$ this leads to significant enhancement of the $\bar{\nu}_e$--flux at HK and hence a slightly better fit. The quality of the fit is not excellent because the energy spectra are distorted enough for the data to ``notice,'' see Fig.~\ref{fig:DUNEvsSK_En}. This result implies that HK, given its huge size, is sensitive to $\tau_3/m_3\lesssim 10^7$~s/eV if the neutrinos are Majorana or if one is interested in the Dirac $\varphi_2$-model (Eq.~\ref{eq:DiracL2}). We have checked this quantitatively and find this is indeed the case.

\section{Concluding Remarks}
\label{sec:conclusions}

We investigated the impact of neutrino decays on the neutronization-burst flux from a supernova. The neutronization-burst flux is less sensitive to the effects of collective oscillations due to the large asymmetry between the rate of $\nu_e$ relative to that of $\bar{\nu}_e$ and hence the reliability of the particle physics information one can extract from it is more robust. Concretely, we considered a model where the heavier neutrino decays into the lightest neutrino and a scalar. In our calculations, we discarded effects related to the masses of the daughter neutrino and scalar. Using the neutronization flux of a $25\,M_\odot $ progenitor, provided by large scale hydrodynamical simulations, we have demonstrated the potential impact in the detected flux of supernova neutrinos in the proposed DUNE and HK detectors, which are expected to be ready by the middle of the next decade. 

Qualitatively, it is straight forward to check that a galactic SN at $10\,{\rm kpc}$ is sensitive to lifetimes $\tau/m \lesssim 10^6$~{s/eV} and is specially sensitive to $\tau/m \sim 10^5$~{s/eV}. There are three distinct effects of the neutrino decay. (i) the overall number of ``visible'' neutrinos can change. This depends on the physics responsible for the neutrino decay. Dirac neutrinos may decay some of the time into right-handed neutrinos -- effectively undetectable -- and some of the time into left-handed daughters -- these can scatter via the charged-current weak interactions with non-zero rates. In a qualitatively distinct scenario, Dirac neutrinos decay some of the time into right-handed antineutrinos (visible) and some of the time into left-handed antineutrinos (invisible). Majorana neutrinos, instead, decay half of the time into ``neutrinos'' and half of the time into ``antineutrinos,'' all of them detectable via charged-current interactions. (ii) the total neutrino energy spectrum softens since one parent neutrino decays into one daughter neutrino with energy that is necessarily smaller than or equal to that of the parent. This effect is more or less pronounced depending on whether there is a ``mismatch'' between the helicities of the parent and daughter neutrinos in the laboratory frame. (iii) the flavor-contents of the different neutrino mass eigenstates are distinct. In particular, $\nu_3$ has a very small $\nu_e$-component -- i.e., $|U_{e3}|^2$ is very small -- while  $\nu_1$ has a large $\nu_e$-component. This means that if one is interested in a $\nu_3\to\nu_1$ decay -- which we encounter in the NMO -- and the detection signal is largest for electron-type neutrinos or antineutrinos, the daughter neutrinos may be much more accessible than the parent neutrinos. The situation is reversed if one is interested in a $\nu_2\to \nu_3$ decay -- which we encounter in the IMO.

We highlighted a few different consequences of the decaying neutrino hypothesis associated to the measurement of the neutronization-burst neutrinos at DUNE and HK. If the neutrino mass ordering is not known, one mass ordering can be ``mimicked'' by the other mass ordering if one allows for the possibility that the heaviest neutrino decays. Results of a concrete hypothesis-test are depicted in Fig.~\ref{fig:ChiSq}. Depending on how far away is the supernova explosion, this mass-ordering-confusion may prevent one from determining the neutrino mass-ordering using only the measurements of the supernova neutronization-burst neutrino flux.  Note that resonant spin-flavor conversions of neutrinos related to a non-zero neutrino magnetic moment can also cause $\nu_e\to\overline{\nu}_e$ conversions in the IMO, thereby resulting in a vanishing neutronization peak. This scenario, however, requires very strong magnetic fields inside the SN $\sim \mathcal{O}(10^{10})\,{\rm G}$ or larger \cite{Ando:2003is,Akhmedov:2003fu}.

On the other hand, if the neutrino mass-ordering is known, measurements of the supernova neutronization-burst neutrino flux allow one to constrain the neutrino lifetime. If data at DUNE are consistent with stable neutrinos, Fig.~\ref{fig:ChiSqTau} reveals that, for a specific model (Dirac neutrinos, $\varphi_0$-scenario), $\tau/m$ values less than $10^5$~s/eV can be safely ruled out if the supernova explosion is not too far away. This sensitivity is far superior to that of solar-system-bound oscillation experiments, by many orders of magnitude ($10^{-3}$ s/eV versus $10^5$~s/eV). In other scenarios (e.g., Dirac neutrinos and the $\varphi_2$-scenario or Majorana neutrinos), introduced here but not explored in great detail in the preceding sections, HK is expected to provide most of the sensitivity. Improvements in the background identification in HK by neutron tagging, which will make it sensitive to the Dirac $\varphi_0$-scenario, will be considered in future extensions of the work.
We have computed the equivalent of Fig.~\ref{fig:ChiSqTau} assuming the neutrinos are Majorana fermions and concentrating on HK data. We find that, for a SN 10~kpc away, HK can rule out $\tau/m$ values less than $10^7$~s/eV, which corresponds to a coupling $|g|\gtrsim 2.3\times 10^{-10}$.

Stronger bounds than the sensitivities discussed here can be extracted from the properties of the cosmic neutrino background, indirectly constrained by cosmic surveys of different types. Precise measurements of the cosmic microwave background from Planck 2015 constrain the neutrino free-streaming length and hence limit the strength of neutrino--neutrino interactions, including those mediated by the light scalars $\varphi$ introduced here. These constraints can be translated
into very strong constraints on the neutrino lifetime $\tau_{\nu} > 4\times 10^{8}\,{\rm s}\,(m_{\nu}/0.05\,\text{eV})^{3}$ for SM neutrinos decaying into lighter neutrinos and dark radiation \cite{Escudero:2019gfk} . These and other cosmological bounds, however, are indirect probes of neutrino decay. We advocate that such bounds are qualitatively distinct and that more direct bounds from ``terrestrial'' experiments complement the more indirect results from cosmic surveys.
 
We emphasized the fact that decaying Majorana and Dirac neutrinos are qualitatively different and potentially easy to distinguish. Qualitatively, the most relevant feature is that, in general, Dirac neutrinos decay either into neutrinos or antineutrinos while Majorana neutrinos decay into both neutrinos and antineutrinos. We showed that, by combining data from DUNE (sensitive to $\nu_e$) and HK (sensitive to $\bar{\nu}_e$) one should be able to distinguish a decaying Dirac neutrino from a Majorana one. The results of a concrete exercise are depicted in Fig.~\ref{fig:ChiSqDvsM}. The complementarity of the two next-generation experiment is quite apparent. We re-emphasize, however, that it is always possible to concoct different models where one cannot distinguish Majorana from Dirac decaying neutrinos, as we discuss in some detail in Sec.~\ref{sec:neutrinodecay}C.

We restricted most of our discussion to one of the models introduced in Sec.~\ref{sec:neutrinodecay}, the $\varphi_0$-model, with $g_{ij}=g_{ji}$, described in Eq.~(\ref{eq:DiracL}). Many of the results discussed here would also apply in the $\varphi_2$-model and in the case where neutrinos are Majorana fermions. In these other scenarios, however, the relative role of DUNE and HK data may be quite distinct. While we concentrated on the decay of neutrinos into other neutrinos and a new scalar particle, there are many other possibilities. In the absence of no new light degrees of freedom the neutrino three-body decays $(\nu_h\to \nu_l \bar{\nu}'_l \nu''_l)$ could lead to some of the same effects discussed here. We plan to return to these decay modes in future work. More very large neutrino experiments, other than DUNE and HK, are expected to be on-line in the latter half of the next decade, including IceCube (ice), KM3Net (salt water), JUNO (liquid scintillator), etc. We did not consider the impact of their data in our many analyses. For example, the large volume and the time resolution of IceCube~\cite{Abbasi:2011ss} have been already exploited in determining the rise time of the $\overline{\nu}_e$ flux during the neutronization burst. The helicity-flipping decays of heavier neutrinos during this epoch could lead to the identification  of a peak at IceCube. However, the uncertainties in the background and the determination of the onset of the signal, as well as the difficulties in the reconstruction of the neutrino energy, will negatively impact the IceCube sensitivity. We plan to return to these and related issues in the future.

\section*{Acknowledgements}
 We would like to thank Edoardo Vitagliano for useful discussions.
 The work of AdG was supported in part by DOE grant \#DE-SC0010143. 
 IMS acknowledges travel support from the Colegio de F\'{\i}sica Fundamental e Interdisciplinaria de las Am\'ericas (COFI). 
MS acknowledges support from the National Science Foundation, Grant PHY-1630782, and to the Heising-Simons Foundation, Grant 2017-228.
Fermilab is operated by the Fermi Research Alliance, LLC under contract No. DE-AC02-07CH11359 with the United States Department of Energy.

\bibliographystyle{kpmod}
\bibliography{DecayNu}

\end{document}